\documentclass[10pt,english]{IEEEtran}
\usepackage[T1]{fontenc}
\usepackage[latin9]{inputenc}
\usepackage{array}
\usepackage{verbatim}
\usepackage{bm}
\usepackage{amsmath}
\usepackage{amssymb}
\usepackage{graphicx}
\usepackage{esint}

\makeatletter

\providecommand{\tabularnewline}{\\}

\makeatother

\usepackage{amsthm}\usepackage{dsfont}\usepackage{array}\usepackage{mathrsfs}\usepackage{cite}\usepackage{comment}\usepackage{mathrsfs}\usepackage{hyperref}

\makeatother

\usepackage{babel}

\usepackage[shortlabels]{enumitem}

\makeatother

\usepackage{babel}
\begin{document}

\title{On the Minimax Capacity Loss \\ Under Sub-Nyquist Universal Sampling}

\author{Yuxin Chen, Andrea J. Goldsmith, and Yonina C. Eldar\thanks{Y. Chen is with the Department of Electrical Engineering, Princeton
University, Princeton, NJ 08544, USA (email: yuxin.chen@princeton.edu).
A. J. Goldsmith is with the Department of Electrical Engineering,
Stanford University, Stanford, CA 94305, USA (email: andrea@ee.stanford.edu).
Y. C. Eldar is with the Department of Electrical Engineering, Technion,
Israel Institute of Technology Haifa, Israel 32000 (email: yonina@ee.technion.ac.il).
The contact author is Y. Chen. This work was supported in part by
the NSF under grant CCF-0939370 and CIS-1320628, the AFOSR under MURI
Grant FA9550-12-1-0215, and BSF Transformative Science Grant 2010505.
It has been presented in part at the IEEE International Symposium
on Information Theory (ISIT) 2013. }}

\maketitle
\theoremstyle{plain}\newtheorem{lem}{\textbf{Lemma}}\newtheorem{theorem}{\textbf{Theorem}}\newtheorem{corollary}{\textbf{Corollary}}\newtheorem{prop}{\textbf{Proposition}}\newtheorem{fct}{\textbf{Fact}}\newtheorem{remark}{\textbf{Remark}}

\theoremstyle{definition}\newtheorem{definition}{\textbf{Definition}}\newtheorem{example}{\textbf{Example}}
\begin{abstract}
This paper investigates the information rate loss in analog channels
when the sampler is designed to operate independent of the instantaneous
channel occupancy. Specifically, a multiband linear time-invariant
Gaussian channel under universal sub-Nyquist sampling is considered.
The entire channel bandwidth is divided into $n$ subbands of equal
bandwidth. At each time only $k$ constant-gain subbands are active,
where the instantaneous subband occupancy is not known at the receiver
and the sampler. We study the information loss through a information
rate loss\emph{ }metric, that is, the gap of achievable rates caused
by the lack of instantaneous subband occupancy information. We characterize
the minimax information rate loss for the sub-Nyquist regime, provided
that the number $n$ of subbands and the SNR are both large. The minimax
limits depend almost solely on the band sparsity factor and the undersampling
factor, modulo some residual terms that vanish as $n$ and SNR grow.
Our results highlight the power of randomized sampling methods (i.e.
the samplers that consist of random periodic modulation and low-pass
filters), which are able to approach the minimax information rate
loss with exponentially high probability. 

\end{abstract}

\textbf{Index Terms}: Channel capacity, minimax sampling, non-asymptotic
random matrix, log-determinant, concentration of spectral measure

\section{Introduction}

The maximum rate of information that can be conveyed through a continuous-time
communication channel is dependent on the sampling technique employed
at the receiver end. In some cutting-edge communication systems, hardware
and cost limitations often preclude sampling at or above the Nyquist
rate, which presents a major bottleneck in transferring wideband and
energy-efficient receiver design paradigms from theory to practice.
Understanding the effects upon capacity of sub-Nyquist sampling is
thus crucial in circumventing this bottleneck.

In many practical scenarios, the occupancy of the communication channel
varies over time. An ideal adaptive sampler can be dynamically optimized
relative to these channel variations. Nevertheless, in most practical
systems, the samplers and the analog-to-digital converters are static
and are designed independent of which subbands are active at any given
time. This has no effect if the sampling rate employed is commensurate
with the maximum bandwidth (or the Nyquist rate) of the channel. However,
in the sub-Nyquist regime, the sampler design significantly impacts
the information rate achievable over a given channel. As was shown
in \cite{ChenEldarGoldsmith2012}, the capacity-maximizing sub-Nyquist
sampling mechanism depends on knowledge of the channel spectrum. When
the subbands available for communications are time-varying and a universal
(static) sub-Nyquist sampler is used, \emph{capacity loss} is typically
incurred and our work characterizes this loss. 

In the present paper, we consider a linear time-invariant (LTI) Gaussian
channel with known channel gain, whereby the entire channel bandwidth
is divided into $n$ subbands of equal bandwidth. At each timeframe,
only a subset of $k$ subbands are active for transmission, but the
spectral occupancy information is not available at either the receiver
or the sampler. The goal is to explore universal (channel-independent)
design of a sub-Nyquist sampling system that is robust vis-a-vis the
uncertainty of instantaneous channel occupancy. In particular, we
aim to understand the resulting loss of information rates between
sampling with and without subband occupancy information in some minimax
sense (as will be detailed in Section \ref{sub:Universal-Sampling}),
and design a sub-Nyquist sampling system under which the information
rate loss can be uniformly controlled and optimized over all possible
channel support.

\subsection{Related Work}

In various scenarios, sampling above the Nyquist rate is not necessary
for preserving signal information in the sense that it generates a
discrete-time sufficient statistic, provided that certain signal structures
are appropriately exploited \cite{ButSte1992,MisEld2011}. Take multiband
signals for example, that reside within several subbands over a wide
spectrum. If the spectral support is known, then the sampling rate
necessary for perfect signal reconstruction is the spectral occupancy,
termed the \textit{Landau rate} \cite{Landau1967}. Such signals admit
perfect recovery when sampled at rates approaching the Landau rate,
assuming appropriately chosen sampling sets (e.g. \cite{HerleyWong1999,VenkataramaniBresler2001}).
Inspired by recent ``compressed sensing'' \cite{CandRomTao06,Don2006,eldar2012compressed}
ideas, spectrum-blind sub-Nyquist samplers have also been developed
for multiband signals \cite{MisEld2010Theory2Practice}. Two of the
most widely used modules employed in the sampler designs are filter
banks and periodic modulation \cite{Papoulis1977,LinVai1998,gardner1986spectral,MisEld2010Theory2Practice}.
These sampling-theoretic works, however, were not based on capacity
as a metric in the sampler design.

On the other hand, the Shannon-Nyquist sampling theorem has frequently
been invoked to investigate the capacity of analog waveform channels
(e.g. \cite{Med2000,Bello1963}). The effects upon capacity of oversampling
have been investigated as well in the presence of quantization \cite{Sha1994,KochLap2010}.
However, none of these works considered the effect of undersampling
upon capacity. Another recent line of work \cite{PelegShamai} investigated
the tradeoff between sparse coding and subsampling in AWGN channels,
but did not consider capacity-achieving input distributions. 

Our recent work \cite{ChenGolEld2010,ChenEldarGoldsmith2012} established
a new framework for investigating the capacity of LTI Gaussian channels
under a broad class of sub-Nyquist sampling strategies, including
filter-bank and modulation-bank sampling and, more generally, time-preserving
sampling. We demonstrated that sampling with a filter bank is sufficient
to approach maximum capacity, assuming that perfect channel state
information (CSI) is available at both the receiver and the transmitter.
In many practical scenarios, however, the active frequency set available
for communications might be changing over time, like in cognitive
radio networks where the spectral subbands available to cognitive
users are varying over time. To the best of our knowledge, no prior
work has investigated, from a capacity perspective, a channel-blind
sub-Nyquist sampling paradigm in the absence of subband occupancy
information. 

Finally, the effect of undersampling has been explored from a source
coding perspective as well. For instance, the fundamental rate-distortion
function of Gaussian sources has been determined under sub-Nyquist
sampling with filtering \cite{kipnis2014distortion}, revealing that
the alias suppressing sampler design achieves the optimal rate distortion
function. For the case where the input source signals are sparse,
the recent work \cite{Narayan2014} characterized the rate-distortion
function under independent and memoryless random sampling. The main
results and techniques presented herein might potentially extend to
these source coding settings to quantify the rate loss caused by spectral-blind
sampling design. 

\nocite{chen2014backing}

\subsection{Main Contributions}

We consider a frequency-flat multiband channel model and the class
of sampling systems with filter banks and periodic modulation. For
this model, our main contributions are summarized as follows.

\begin{itemize}\itemsep0.5em \item We derive a lower bound (Theorem
\ref{Theorem-Minimax-Det-Upper-Bound}) on the minimax sampled information
rate loss (defined in Section \ref{sec:Problem-Formulation-Minimax})
incurred due to the lack of channel occupancy information, under super-Nyquist
universal sampling. This minimax lower limit depends almost only on
the band sparsity factor and the undersampling factor, modulo some
residual terms that vanish when SNR and $n$ increase. 

\item We characterize in Theorem \ref{theorem-Minimax-Sampler-BWLimited-Achievability-Oversampling}
the sampled information rate loss under a class of sampling systems
with periodic modulation and low-pass filters with passband $\left[0,W/n\right]$,
when the Fourier coefficients of the modulation waveforms are generated
in an i.i.d. Gaussian fashion (termed \emph{Gaussian sampling}). We
demonstrate that with exponentially high probability, the resulting
sampled information rate loss matches the lower bound given in Theorem
\ref{Theorem-Minimax-Det-Upper-Bound} uniformly over all possible
subband occupancy. This implies that random sampling strategies are
minimax-optimal in terms of a universal sampling design.

\item The power of random sampling arises due to sharp concentration
of spectral measures of large random matrices \cite{GuionnetZeitouni2000}.
To establish Theorem \ref{theorem-Minimax-Sampler-BWLimited-Achievability-Oversampling},
we derive measure concentration of several log-determinant functions
for i.i.d. Gaussian ensembles, which might be of independent interest
for other works involving log-determinant metrics. 

\end{itemize}

\subsection{Organization}

The remainder of this paper is organized as follows. In Section \ref{sec:Problem-Formulation-Minimax}
we introduce our system model of multiband Gaussian channels. A metric
called sampled information rate loss, and a minimax sampler, are defined
with respect to sampled channel capacity. We then determine in Section
\ref{sec:Minimax-Capacity-Regret} the minimax information rate loss.
Specifically, we develop lower bounds on the minimax information rate
loss in Section \ref{sub:The-Converse-Minimax}. The achievability
is treated in Section \ref{sub:Achievability-Super-Landau-Minimax}.
Besides, we derive measure concentration of several log-determinant
functions in Section \ref{sub:Concentration-of-Log-Determinant}.
Section \ref{sub:Implications-Minimax} summarizes the key observation
and implications from our results. Section \ref{sec:Conclusion-Minimax}
closes the paper with a short summary of our findings and potential
future directions.

\subsection{Notation}

Denote by $\mathcal{H}(\beta):=-\beta\log\beta-(1-\beta)\log(1-\beta)$
the binary entropy function. The standard notation $f(n)=\mathcal{O}\left(g(n)\right)$
means there exists a constant $c>0$ such that $\left|f(n)\right|\leq c|g(n)|$,
$f(n)=\Theta\left(g(n)\right)$ means there exist constants $c_{1},c_{2}>0$
such that $c_{1}|g(n)|\leq|f(n)|\leq c_{2}|g(n)|$, $f(n)=\omega\left(g(n)\right)$
means that $\lim_{n\rightarrow\infty}\frac{g(n)}{f(n)}=0$, and $f(n)=o\left(g(n)\right)$
indicates that $\lim_{n\rightarrow\infty}\frac{f(n)}{g(n)}=0$. For
a matrix $\boldsymbol{A}$, we use $\boldsymbol{A}_{i*}$ and $\boldsymbol{A}_{*i}$
to denote the $i$th row and $i$th column of $\boldsymbol{A}$, respectively.
We let $[n]$ denote the set $\left\{ 1,2,\cdots,n\right\} $, and
write ${[n] \choose k}$ for the set of all $k$-element subsets of
$\left\{ 1,2,\cdots,n\right\} $. We also use $\text{card}\left(\mathcal{A}\right)$
to denote the cardinality of a set $\mathcal{A}$. Let $\boldsymbol{W}$
be a $p\times p$ random matrix that can be expressed as $\boldsymbol{W}=\Sigma_{i=1}^{n}\boldsymbol{Z}_{i}\boldsymbol{Z}_{i}^{\top}$,
where $\boldsymbol{Z}_{i}\sim\mathcal{N}\left(0,\boldsymbol{\Sigma}\right)$
are jointly independent Gaussian vectors. Then $\boldsymbol{W}$ is
said to have a central Wishart distribution with $n$ degrees of freedom
and scale matrix $\boldsymbol{\Sigma}$, denoted by $\boldsymbol{W}\sim\mathcal{W}_{p}(n,\boldsymbol{\Sigma})$.
Our notation is summarized in Table \ref{tab:Summary-of-Notation-Nonuniform}. 

\begin{table}
\caption{\label{tab:Summary-of-Notation-Nonuniform}Summary of Notation and
Parameters}

\centering{}%
\begin{tabular}{l>{\raggedright}p{0.77\columnwidth}}
$\mathcal{H}(x)$  & binary entropy function, i.e. $\mathcal{H}(x)=-x\log x-(1-x)\log(1-x)$ \tabularnewline
$h(t)$,$H(f)$  & impulse response, and frequency response of the LTI analog channel\tabularnewline
$\mathcal{S}_{\eta}(f)$ & power spectral density of the additive Gaussian noise $\eta(t)$\tabularnewline
$f_{\text{s}}$, $T_{\text{s}}$ & aggregate sampling rate ($f_{\text{s}}=\frac{m}{n}W$), and the corresponding
sampling interval ($T_{\text{s}}=1/f_{\text{s}}$)\tabularnewline
$W$, $W_{0}$ & channel bandwidth, size of instantaneous channel support\tabularnewline
$n,m,k$ & number of subbands, number of sampling branches, number of subbands
being simultaneously active\tabularnewline
$\alpha=m/n$ & undersampling factor\tabularnewline
$\beta=k/n$ & sparsity factor\tabularnewline
$\boldsymbol{Q}$, $\boldsymbol{Q}^{\text{w}}$ & sampling matrix, whitened sampling matrix ($\boldsymbol{Q}^{\text{w}}=\left(\boldsymbol{Q}\boldsymbol{Q}^{*}\right)^{-\frac{1}{2}}\boldsymbol{Q}$)\tabularnewline
$L_{\boldsymbol{s}}^{\boldsymbol{Q}}$ & information rate loss associated with a sampling matrix $\boldsymbol{Q}$
given state $\boldsymbol{s}$\tabularnewline
$\boldsymbol{A}_{i*}$, $\boldsymbol{A}_{*i}$ & $i$th row of $\boldsymbol{A}$, $i$th column of $\boldsymbol{A}$\tabularnewline
$\text{card}\left(\mathcal{A}\right)$ & cardinality of a set $\mathcal{A}$\tabularnewline
$[n]$ & $[n]:=\left\{ 1,2,\cdots,n\right\} $\tabularnewline
${[n] \choose k}$  & set of all $k$-element subsets of $[n]$\tabularnewline
$\mathcal{W}_{p}\left(n,\boldsymbol{\Sigma}\right)$ & $p$-dimensional central Wishart distribution with $n$ degrees of
freedom and scale matrix $\boldsymbol{\Sigma}$ \tabularnewline
\end{tabular}
\end{table}

\section{Problem Formulation and Preliminaries\label{sec:Problem-Formulation-Minimax}}

\subsection{Compound Multiband Channel\label{sub:Compound-Multiband-Channel}}

We consider a multiband Gaussian channel of total bandwidth $W$,
and it is divided into $n$ continuous subbands\footnote{Note that in practice, $n$ is typically a large number. For instance,
the number of subcarriers ranges from 128 to 2048 in LTE \cite{ghosh2010fundamentals,LTEWhitePaper}.} each of bandwidth $W/n$. A state $\boldsymbol{s}\in{[n] \choose k}$
is generated, which dictates the channel support. For ease of presentation,
the present work focuses on the frequency-flat channel model, which
suffices to capture the essence of our findings. Specifically, given
a state $\boldsymbol{s}$, the channel is assumed to be an LTI filter
with impulse response $h_{\boldsymbol{s}}(t)$ and frequency response
\begin{equation}
H_{\boldsymbol{s}}(f)=\begin{cases}
H,\quad & \text{if }f\text{ }\text{lies within subbands at indices from }\boldsymbol{s},\\
0, & \text{else}.
\end{cases}\label{eq:ChannelGain}
\end{equation}
A transmit signal $x(t)$ with a power constraint $P$ is passed through
this multiband channel, yielding a channel output
\begin{equation}
r_{\boldsymbol{s}}(t)=h_{\boldsymbol{s}}(t)*x(t)+\eta(t),\label{eq:ChannelModel-1}
\end{equation}
where $\eta(t)$ is stationary zero-mean Gaussian noise with power
spectral density $\mathcal{S}_{\eta}\left(f\right)\equiv1$.  It is
assumed throughout that the knowledge of $H$ and $\mathcal{S}_{\eta}$
are available at both the transmitter and the receiver, while the
state $\boldsymbol{s}$ is known only at the transmitter. The results
derived herein can be extended to more general frequency selective
channels with optimal power control. These extensions are described
in more details in Section \ref{sub:Extension} and derived in \cite{chen2014Thesis}. 

\subsection{Sampled Channel Capacity\label{sub:Sampled-Channel-Capacity-Minimax}}

We aim to design a sampler that operates below the Nyquist rate (i.e.
the channel bandwidth $W$). In particular, the present work focuses
on the class of filter-bank and modulation-bank sampling systems,
which subsumes the most widely used sampling mechanisms in practice.

\subsubsection{Sampling System and Channel Capacity}

\begin{figure}
\begin{centering}
\begin{tabular}{c}
\includegraphics[scale=0.35]{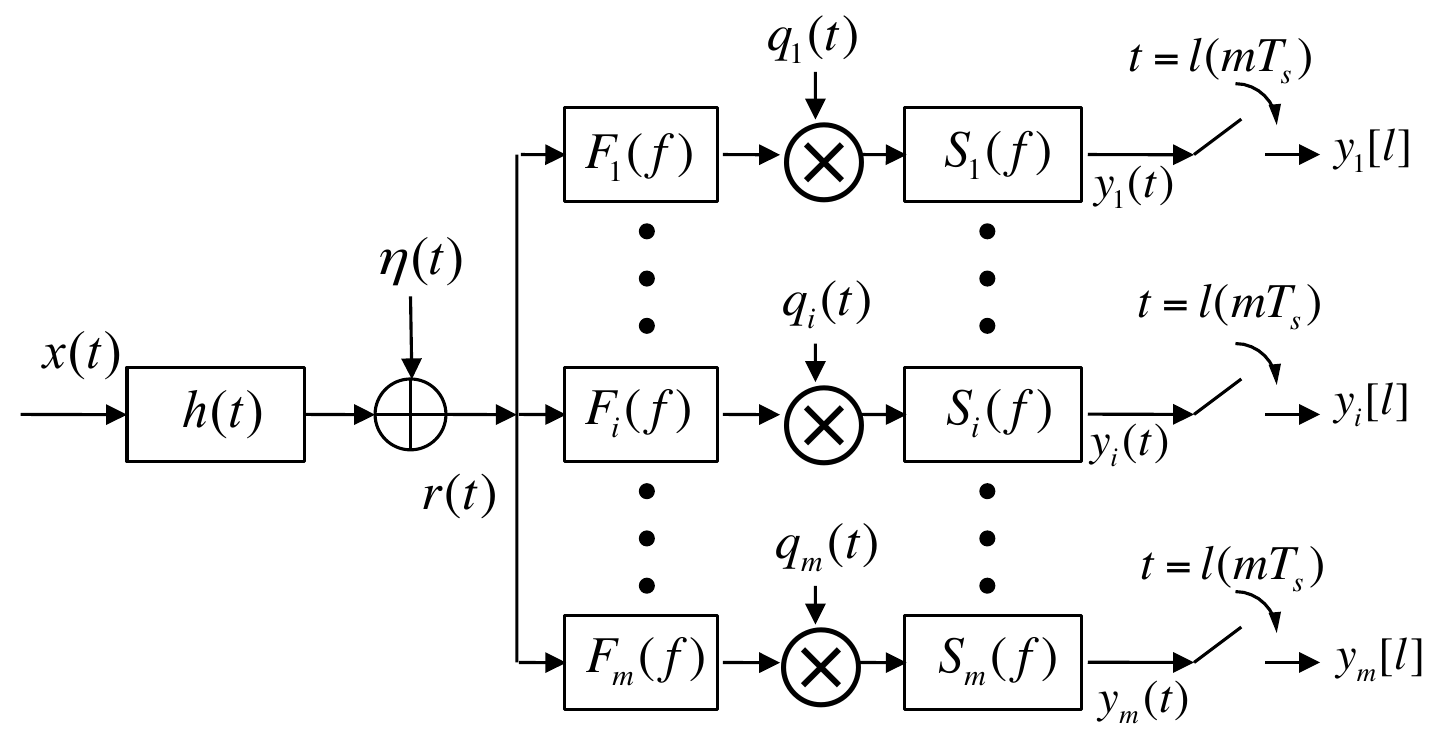}\tabularnewline
(a)\tabularnewline
\includegraphics[scale=0.35]{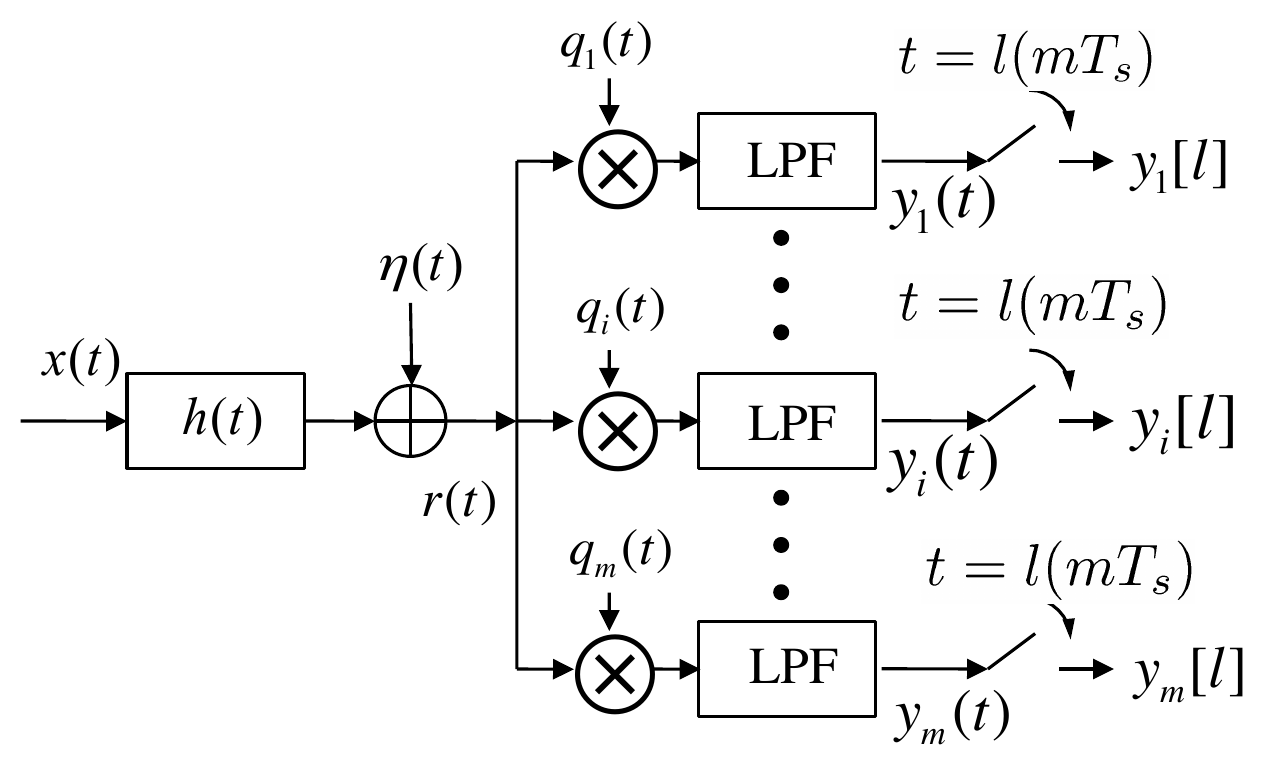}\tabularnewline
(b)\tabularnewline
\end{tabular}
\par\end{centering}

\caption{\label{fig:SamplingModulationFilter}(a) Sampling with modulation
and filter banks: the channel output $r(t)$ is passed through $m$
branches, each consisting of a pre-modulation filter, a periodic modulator
and a post-modulation filter followed by a uniform sampler with sampling
rate $W/n$. (b) Sampling with a bank of modulators and low-pass filters:
the channel output is passed through $m$ branches, each consisting
of a modulator with modulation waveform $q_{i}(t)$ and a low-pass
filter of pass band $\left[0,W/n\right]$ followed by a uniform sampler
at rate $W/n$.}
\end{figure}

We consider the class of sampling systems that consist of a combination
of filter banks and periodic modulation, as illustrated in Fig. \ref{fig:SamplingModulationFilter}(a).
Specifically, the sampling system comprises $m$ branches, where at
the $i$th branch, the channel output is passed through a pre-modulation
LTI filter $F_{i}\left(f\right)$, modulated by a periodic waveform
$q_{i}\left(t\right)$ of period $T_{q}=\left(W/n\right)^{-1}$, and
then passed through a post-modulation LTI filter $S_{i}\left(f\right)$
followed by uniform sampling at rate $W/n$. The aggregate sampling
rate is $f_{\mathrm{s}}=\frac{m}{n}W$. When specialized to the frequency-flat
channels, it is natural to concentrate on the case where $F_{i}\left(f\right)$
and $S_{i}\left(f\right)$ are both flat\footnote{Our main results and analytical tools can be extended to more general
frequency-varying periodic sampling systems without difficulty. Interested
readers are referred to \cite{chen2014Thesis} for details.} within each subband $\left[\frac{lW}{n},\frac{\left(l+1\right)W}{n}\right)$
($l\in\mathbb{Z}$).

Since the modulation waveform $q_{i}(t)$ is periodic, its Fourier
transform can be represented by a weighted $\delta$-train, namely,
\begin{equation}
\mathcal{F}\left(q_{i}(t)\right)=\sum_{l=-\infty}^{\infty}\hat{q}_{i,l}\delta\left(f+lW/n\right)
\end{equation}
for some sequence $\left\{ \hat{q}_{i,l}\right\} _{l\in\mathbb{Z}}$.
This modulation operation scrambles the spectral content of the channel
input $X\left(f\right)$. As can be seen, the signal after post-modulation
filtering (i.e. $y_{i}(t)$ in Fig. \ref{fig:SamplingModulationFilter}(a))
has Fourier response
\begin{equation}
\sum_{l=0}^{n}\hat{q}_{i,l}S_{i}\left(f\right)F_{i}\left(f+l\frac{W}{n}\right)X\left(f+l\frac{W}{n}\right),\quad\forall f.
\end{equation}
Due to aliasing, the final sampling output (i.e. $y_{i}[n]$ in Fig.
\ref{fig:SamplingModulationFilter}(a)) is tantamount to a signal
of Fourier responses
\begin{align}
 & \sum_{\tau=-\infty}^{\infty}\sum_{l=0}^{n-1}\hat{q}_{i,l}S_{i}\left(f+\tau\frac{W}{n}\right)F_{i}\left(f+\left(\tau+l\right)\frac{W}{n}\right)\nonumber \\
 & \quad\quad\quad\quad\cdot X\left(f+\left(\tau+l\right)\frac{W}{n}\right),\quad\quad0\leq f<\frac{W}{n}.\label{eq:DefnQ}
\end{align}
Since $F_{i}\left(f\right)$, $S_{i}\left(f\right)$ are piecewise
flat, one can write (\ref{eq:DefnQ}) as 
\begin{equation}
\sum_{l=1}^{n}\boldsymbol{Q}_{i,l}X\left(f+\left(l-1\right)\frac{W}{n}\right),\quad0\leq f<\frac{W}{n}\label{eq:SamplingOutput}
\end{equation}
for some sequence $\left\{ \boldsymbol{Q}_{i,l}\right\} _{1\leq i\leq m,1\leq l\leq n}$.
As a result, one can use an $m\times n$ matrix $\boldsymbol{Q}=\left[\boldsymbol{Q}_{i,l}\right]_{1\leq i\leq m,1\leq l\leq n}$\emph{,}
to represent the sampling system, termed a \emph{sampling coefficient
matrix}. 

On the other hand, for any given $\boldsymbol{Q}\in\mathbb{C}^{m\times n}$,
there exists a sampling system such that the Fourier response of its
sampled output obeys (\ref{eq:SamplingOutput}). This can be realized
via the $m$-branch sampling system illustrated in Fig. \ref{fig:SamplingModulationFilter}(b).
In the $i$th branch, the channel output is modulated by a periodic
waveform $q_{i}(t)$ with Fourier response
\[
\mathcal{F}\left(q_{i}(t)\right)=\sum_{l=1}^{n}\boldsymbol{Q}_{i,l}\delta\left(f+\left(l-1\right)W/n\right),
\]
passed through a low-pass filter with pass band $[0,W/n]$, and then
uniformly sampled at rate $W/n$. In the current paper, a sampling
system within this class is said to be \emph{Gaussian sampling} if
the entries of $\boldsymbol{Q}$ are i.i.d. Gaussian random variables.
It turns out that Gaussian sampling structures suffice to achieve
overall robustness in terms of sampled information rate loss, as will
be seen in Section \ref{sec:Minimax-Capacity-Regret}. 

\begin{figure*}[tbph]
\begin{centering}
\textsf{}%
\begin{tabular}{cc}
\textsf{\includegraphics[width=0.49\textwidth]{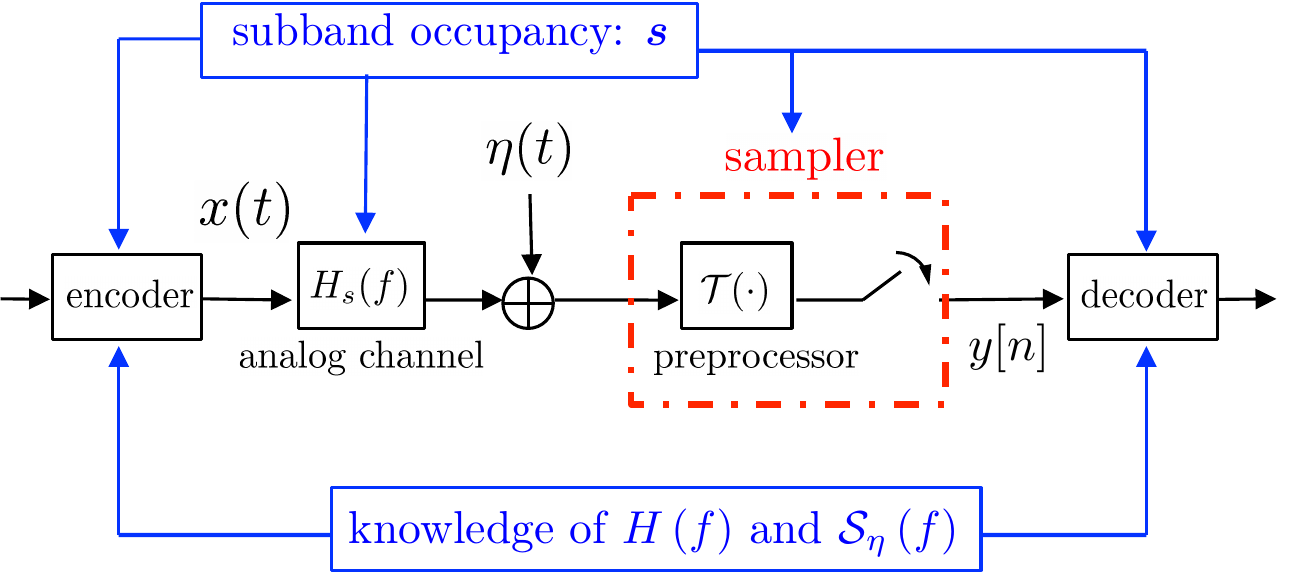}} & \textsf{\includegraphics[width=0.49\textwidth]{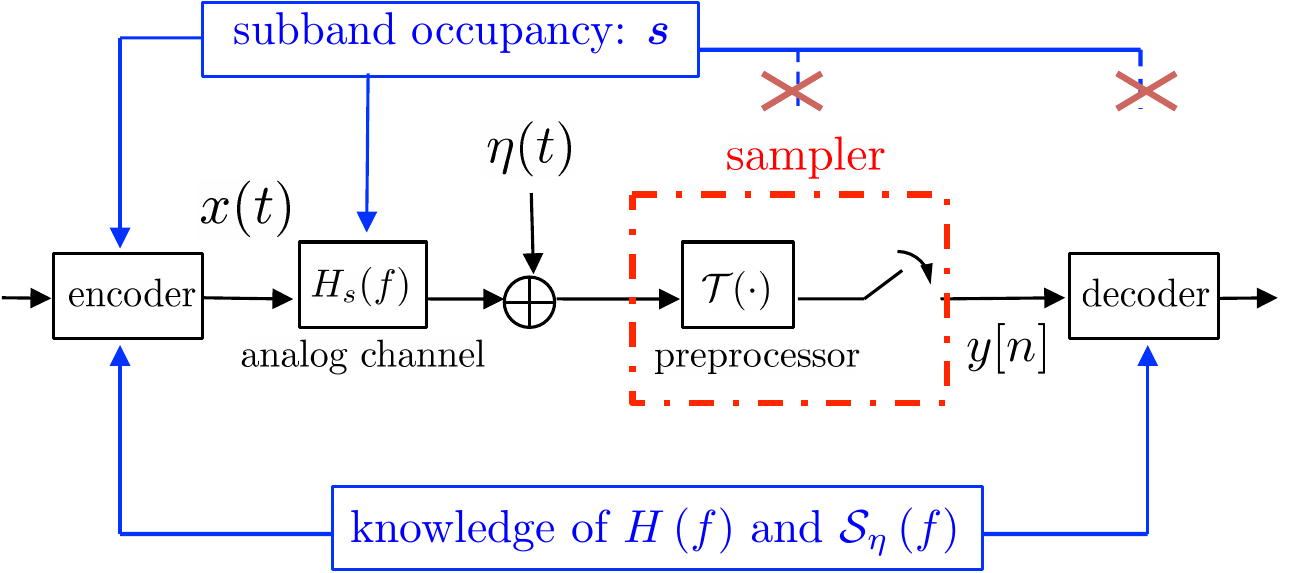}}\tabularnewline
(a) sampling with subband occupancy information & (b) sampling without subband occupancy information\tabularnewline
\end{tabular}
\par\end{centering}

\caption{\label{fig:ProblemFormulation-Minimax}Sampling with subband occupancy
information vs. sampling without subband occupancy information (universal
sampling). }
\end{figure*}

\subsection{Universal Sampling\label{sub:Universal-Sampling}}

As was shown in \cite{ChenEldarGoldsmith2012}, the optimal sampling
mechanism for a given LTI channel with perfect CSI extracts out the
frequency set with the highest SNR and hence suppresses aliasing.
Such an alias-suppressing sampler may result in a very low capacity
for some channel support. In this paper, we desire a sampler that
operates independent of the instantaneous subband occupancy, and our
objective is to design a single linear sampling system that incurs
minimal information rate loss across all possible channel occupancy.
In particular, the information rate loss we consider is the gap between
the capacity under sampling with and without spectral occupancy (i.e.
Fig. \ref{fig:ProblemFormulation-Minimax}(a) vs. Fig. \ref{fig:ProblemFormulation-Minimax}(b)).

\subsubsection{Sampled Information Rate Loss}

For notational convenience, define the \emph{undersampling factor}
and the \emph{sparsity factor} as 
\begin{equation}
\begin{cases}
\alpha & :=m/n,\\
\beta & :=k/n,
\end{cases}\label{eq:Defn-alpha-beta}
\end{equation}
respectively. It will be assumed throughout that $\alpha,\beta\in(0,1)$
are some constants independent of $n$.

Our prior work \cite{ChenEldarGoldsmith2012} reveals that for any
given state $\boldsymbol{s}$ and sampling rate $f_{\mathrm{s}}=\alpha W$,
the capacity under channel-optimized sampling is given by
\begin{align}
C_{\boldsymbol{s}} & =\frac{W}{2n}\min\left\{ k,m\right\} \log\left(1+\frac{P}{\min\left\{ \alpha,\beta\right\} W}\frac{\left|H\right|^{2}}{\mathcal{S}_{\eta}}\right)\nonumber \\
 & =\frac{W}{2}\min\left\{ \alpha,\beta\right\} \log\left(1+\mathsf{SNR}\right),\label{eq:MatrixFormCs}
\end{align}
where we set
\begin{equation}
\mathsf{SNR}:=\frac{P}{\min\left\{ \alpha,\beta\right\} W}\frac{\left|H\right|^{2}}{\mathcal{S}_{\eta}}.\label{eq:SNR-defn}
\end{equation}
In addition, the channel capacity under the aforementioned filter-bank
and modulation-bank sampling has also been derived \cite[Theorem 5]{ChenEldarGoldsmith2012}.
When specialized to the frequency-flat channel model under uniform
power allocation, the achievable rate at a given state $\bm{s}$ without
subband occupancy information is given by
\begin{align}
C_{\boldsymbol{s}}^{\boldsymbol{Q}} & =\text{ }\frac{W}{2n}\log\det\left(\boldsymbol{I}_{m}+\mathsf{SNR}\cdot\left(\boldsymbol{Q}\boldsymbol{Q}^{*}\right)^{-\frac{1}{2}}\boldsymbol{Q}_{\boldsymbol{s}}\boldsymbol{Q}_{\boldsymbol{s}}^{*}\left(\boldsymbol{Q}\boldsymbol{Q}^{*}\right)^{-\frac{1}{2}}\right)\nonumber \\
 & :=\frac{W}{2n}\log\det\left(\boldsymbol{I}_{m}+\mathsf{SNR}\cdot\boldsymbol{Q}_{\boldsymbol{s}}^{\mathrm{w}}\boldsymbol{Q}_{\boldsymbol{s}}^{\mathrm{w}*}\right).\label{eq:CapacityPeriodicSampling}
\end{align}
Here, we let $\boldsymbol{A}_{\boldsymbol{s}}$ represent the submatrix
of $\boldsymbol{A}$ consisting of the columns at indices of $\boldsymbol{s}$,
and $\boldsymbol{Q}^{\mathrm{w}}:=\left(\boldsymbol{Q}\boldsymbol{Q}^{*}\right)^{-\frac{1}{2}}\boldsymbol{Q}$
the prewhitened sampling coefficient matrix. Note that $\boldsymbol{Q}^{\mathrm{w}}\boldsymbol{Q}^{\mathrm{w}*}:=\boldsymbol{I}_{m}.$ 

For any sampling system with a sampling coefficient matrix $\boldsymbol{Q}$,
we define the \emph{sampled information rate loss} for each state
$\boldsymbol{s}$ as
\begin{equation}
L_{\boldsymbol{s}}^{\boldsymbol{Q}}:=C_{\boldsymbol{s}}-C_{\boldsymbol{s}}^{\boldsymbol{Q}}.\label{eq:DefnCapLoss}
\end{equation}
This metric quantifies the information rate loss of universal sampling
due to the lack of subband occupancy information, i.e. the gap of
achievable rates under the sampler in Fig. \ref{fig:ProblemFormulation-Minimax}(a)
relative to the sampler in Fig. \ref{fig:ProblemFormulation-Minimax}(b).

\subsubsection{Minimax Sampler}

We aim to design a sampler that minimizes the loss function in some
overall sense. Let $L$ represent the \emph{minimax information rate
loss}, that is,
\begin{equation}
L:=\inf_{\boldsymbol{Q}}\max_{\boldsymbol{s}\in{[n] \choose k}}L_{\boldsymbol{s}}^{\boldsymbol{Q}}.\label{eq:minimax-loss-defn}
\end{equation}
A sampling system associated with a sampling coefficient matrix $\boldsymbol{M}$
is then called a \emph{minimax sampler} if it satisfies
\begin{equation}
\max_{\boldsymbol{s}\in{[n] \choose k}}L_{\boldsymbol{s}}^{\boldsymbol{M}}=L.\label{eq:DefnMinimaxLoss}
\end{equation}

\begin{figure}
\begin{centering}
\emph{\includegraphics[scale=0.38]{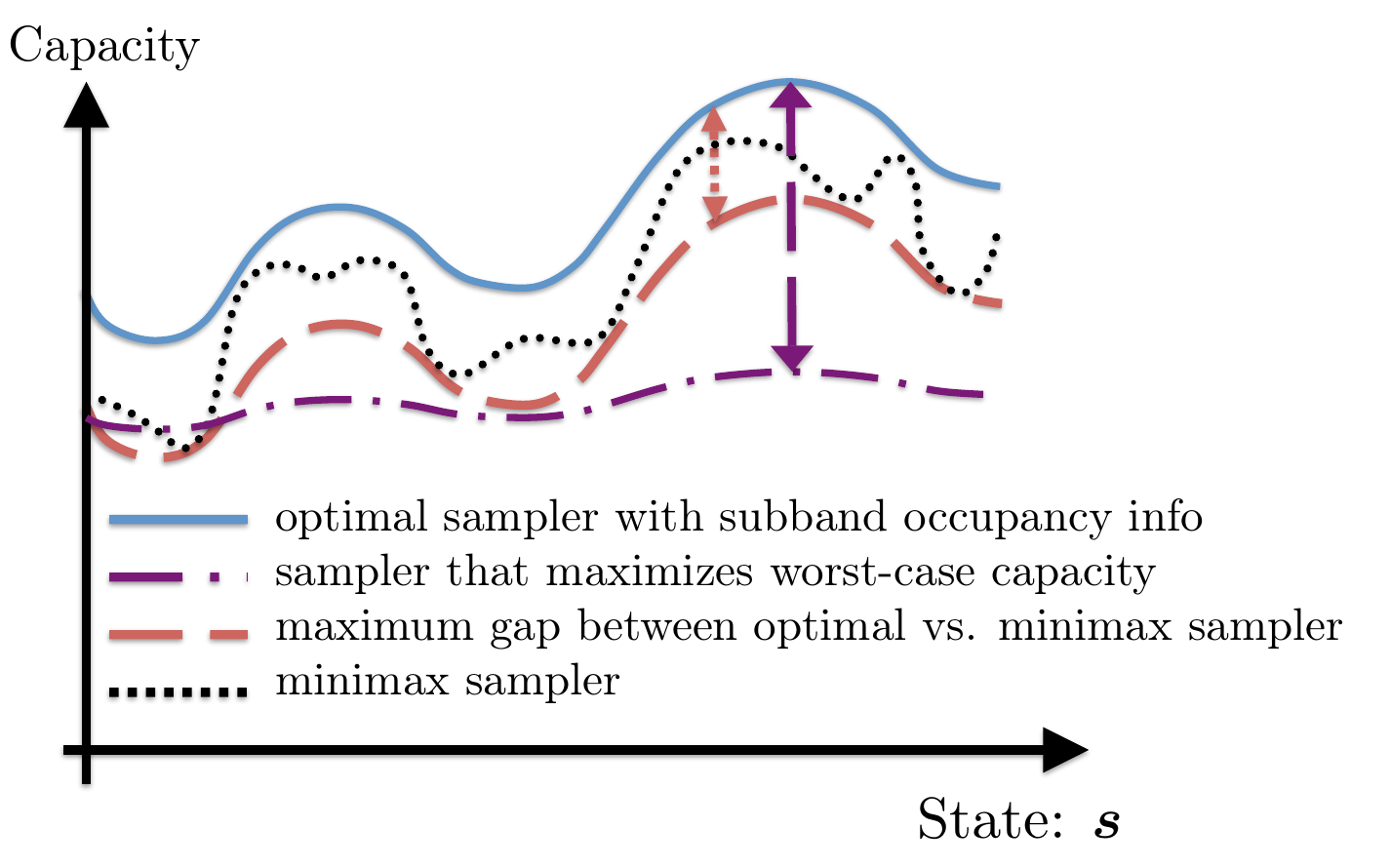}}
\par\end{centering}

\caption{\label{fig:CapacityLoss}Minimax sampler vs. the sampler that maximizes
worst-case capacity. The blue solid line represents the capacity under
channel-optimized sampling with subband occupancy information, the
black dotted line represents the capacity achieved by minimax sampler,
the orange dashed illustrates the maximum capacity minus the minimax
information rate loss, while the purple dashed line corresponds to
maximum worst-case capacity. }
\end{figure}

The minimax criterion is of interest for designing a sampler robust
against all possible channel occupancy situations, that is, we expect
the resulting sampled channel capacity to be within a minimal gap
relative to maximum capacity in a uniform manner. Note that the minimax
sampler is in general different from the one that maximizes the lowest
capacity among all states (worst-case capacity). While the latter
guarantees an optimal worst-case capacity that can be achieved regardless
of which channel is realized, it may result in significant information
rate loss in many other states, as illustrated in Fig. \ref{fig:CapacityLoss}. 

\section{Minimax Sampled Information Rate Loss\label{sec:Minimax-Capacity-Regret}}

The minimax sampled information rate loss problem boils down to minimizing
$\max_{\boldsymbol{s}}L_{\boldsymbol{s}}^{\boldsymbol{Q}}$ over all
sampling coefficient functions $\boldsymbol{Q}$. In general, this
problem is non-convex in $\boldsymbol{Q}$, and hence it is computationally
intractable to find the optimal sampler by solving a numerical optimization
program. Fortunately, for the entire sub-Nyquist regime, the minimax
sampled information rate loss can be quantified reasonably well at
moderate-to-high SNR, and can be well approached by a sampler generated
in a random fashion. 

Our main results are summarized in the following theorem.

\begin{theorem}\label{theorem-Minimax-Final} Suppose that $0<\alpha,\beta<1$.
Define
\begin{equation}
\Lambda:=\min\left\{ \beta,\alpha\right\} \log\left(1+\frac{1}{\small\mathrm{SNR}}\right);\label{eq:Delta_alpha_beta}
\end{equation}
\begin{equation}
\Psi_{1}:=\min\left\{ \left\lceil \frac{\beta}{\alpha-\beta}\right\rceil \frac{1}{\small\mathrm{SNR}},\left(1+\beta\right)\frac{1}{\sqrt{\small\mathrm{SNR}}}\right\} ;\label{eq:Defn-Pi-1}
\end{equation}
\begin{equation}
\Psi_{2}:=\min\left\{ \left\lceil \frac{1-\alpha}{\beta-\alpha}\right\rceil \frac{1}{\small\mathsf{SNR}},\left(1+\alpha\right)\frac{1}{\sqrt{\mathrm{SNR}}}\right\} .\label{eq:Defn-Pi-2}
\end{equation}
(a) If $\beta<\alpha$ and $\alpha+\beta<1$, then
\begin{equation}
L=\frac{W}{2}\left\{ \mathcal{H}\left(\beta\right)-\alpha\mathcal{H}\left(\frac{\beta}{\alpha}\right)+\Delta_{1}\right\} ;\label{eq:MinimaxCapacityGapSuperLandauFinal}
\end{equation}
(b) If $\beta>\alpha$, then
\begin{equation}
L=\frac{W}{2}\left\{ \mathcal{H}\left(\alpha\right)-\beta\mathcal{H}\left(\frac{\alpha}{\beta}\right)+\Delta_{2}\right\} ;\label{eq:MinimaxCapacityGapSuperLandauFinal-2}
\end{equation}
(c) If $\beta=\alpha$ or if $\beta<\alpha$ and $\alpha+\beta\geq1$,
then
\begin{equation}
L=\frac{W}{2}\left\{ \mathcal{H}\left(\beta\right)-\alpha\mathcal{H}\left(\frac{\beta}{\alpha}\right)+\Delta_{3}\right\} .\label{eq:MinimaxCapacityGapSuperLandauFinal-3}
\end{equation}
Here, $\Delta_{1}\sim\Delta_{3}$ are some residual terms obeying
\begin{align*}
\Lambda-\Psi_{1}-\frac{\log\left(n+1\right)}{n} & \leq\Delta_{1}\leq\Lambda+\frac{c_{1}\log n}{n^{1/3}},\\
\Lambda-\Psi_{2}-\frac{\log\left(n+1\right)}{n} & \leq\Delta_{2}\leq\Lambda+\frac{c_{2}\log n}{n^{1/3}},\\
\Lambda-\Psi_{1}-\frac{\log\left(n+1\right)}{n} & \leq\Delta_{3}\leq\Lambda+\frac{c_{3}\mathsf{SNR}^{\frac{1}{3}}\log n}{n^{1/3}},
\end{align*}
and $c_{1}\sim c_{3}$ are some universal constants independent of
$n$ and $\mathsf{SNR}$.\end{theorem}

\begin{remark}Note that $\mathcal{H}(\cdot)$ denotes the binary
entropy function. Its appearance is due to the fact that it is a tight
estimate of the rate function of binomial coefficients. \end{remark}

Theorem \ref{theorem-Minimax-Final} provides a tight characterization
of the minimax sampled information rate loss relative to the capacity
under channel-optimized sampling. The minimax limits per unit bandwidth
are given by
\begin{equation}
L\approx\begin{cases}
\frac{1}{2}\mathcal{H}\left(\beta\right)-\frac{1}{2}\alpha\mathcal{H}\left(\frac{\beta}{\alpha}\right),\quad & \text{if }\mbox{\ensuremath{\alpha\geq\beta}},\\
\frac{1}{2}\mathcal{H}\left(\alpha\right)-\frac{1}{2}\beta\mathcal{H}\left(\frac{\alpha}{\beta}\right),\quad & \text{if }\mbox{\ensuremath{\alpha\leq\beta}},
\end{cases}\label{eq:first-order}
\end{equation}
modulo some residual terms. For $3/4$ of the sub-Nyquist regime,
the residuals are at most of order $\mathcal{O}\left(\frac{\log n}{n^{1/3}}+\frac{1}{\small\mathrm{SNR}}\right)$,
which are negligible at high SNR and when the number $n$ of subbands
is large. For another $1/4$ of the sub-Nyquist regime, our results
are tight to within a gap $\mathcal{O}\left(\frac{\small\mathsf{SNR}^{1/3}\log n}{n^{1/3}}+\frac{1}{\small\mathrm{SNR}}\right)$,
which will vanish if\footnote{This is a practically common situation. For instance, in the LTE communication
systems, the median-to-high SNR for urban macrocells is typically
between $10\sim20$dB, while the number of sub-carriers is around
$128\sim2048$ \cite[Chapter 26]{sesia2009lte}. } $\frac{\mathrm{SNR}}{n}=o\left(1\right)$. For the special Landau-rate
sampling case (i.e. $\alpha=\beta$), our bounds are accurate up to
some gap $\mathcal{O}\left(\frac{\small\mathrm{SNR}^{1/3}\log n}{n^{1/3}}+\frac{1}{\sqrt{\small\mathrm{SNR}}}\right)$.
We remark, however, that the factor $\frac{\small\mathsf{SNR}^{1/3}\log n}{n^{1/3}}$
is not an optimal order and might be refined by other techniques. 

The proof of Theorem \ref{theorem-Minimax-Final} involves the verification
of two parts: a converse part that provides a  lower bound on the
minimax sampled information rate loss, and an achievability part that
provides a sampling scheme to approach this bound. As we show, the
class of sampling systems with random periodic modulation followed
by low-pass filters, as illustrated in Fig. \ref{fig:SamplingModulationFilter}(b),
is sufficient to approach the minimax sampled loss.

\subsection{Lower Bound on the Minimax Information Rate Loss \label{sub:The-Converse-Minimax}}

We need to demonstrate that the minimax sampled information rate loss
under any channel-independent sampler cannot be lower than (\ref{eq:MinimaxCapacityGapSuperLandauFinal})-(\ref{eq:MinimaxCapacityGapSuperLandauFinal-3})
in respective regimes. This is given by the theorem below.

\begin{theorem}\label{Theorem-Minimax-Det-Upper-Bound}(1) If $\beta\leq\alpha\leq1$,
then
\begin{align}
 & L\geq\frac{W}{2}\left\{ \mathcal{H}\left(\beta\right)-\alpha\mathcal{H}\left(\frac{\beta}{\alpha}\right)+\Lambda-\frac{\log\left(n+1\right)}{n}-\right.\nonumber \\
 & \footnotesize\left.\min\left\{ \left\lceil \frac{\beta}{\alpha-\beta}\right\rceil \log\left(1+\frac{1}{\small\mathrm{SNR}}\right)\right\} ,\left(1+\beta\right)\log\left(1+\frac{1}{\small\mathrm{SNR}^{\frac{1}{2}}}\right)\right\} .\label{eq:MinimaxCapacityGapLowerBoundLargeN}
\end{align}
(2) If $0<\alpha<\beta$, then 
\begin{align}
 & L\geq\frac{W}{2}\left\{ \mathcal{H}\left(\alpha\right)-\beta\mathcal{H}\left(\frac{\alpha}{\beta}\right)+\Lambda-\frac{\log\left(n+1\right)}{n}-\right.\nonumber \\
 & \quad\footnotesize\min\left\{ \left\lceil \frac{1-\alpha}{\beta-\alpha}\right\rceil \log\left(1+\frac{1}{\small\mathsf{SNR}}\right),\left(1+\alpha\right)\log\left(1+\frac{1}{\small\mathrm{SNR}^{\frac{1}{2}}}\right)\right\} 
\end{align}
Here, $\Lambda$ is defined in (\ref{eq:Delta_alpha_beta}).\end{theorem}

For the entire sub-Nyquist regime, the lower bounds we derive are
tantamount to some constants dependent only on $\alpha$ and $\beta$,
except for some residual terms that vanish when the number $n$ of
subbands and the SNR tend to infinity. More precisely, when $\alpha\neq\beta$,
one has $\Lambda,\Psi=\mathcal{O}\left(\frac{1}{\small\mathsf{SNR}}\right)$,
and hence these residuals are at most the order of $\mathcal{O}\left(\frac{1}{\small\mathsf{SNR}}+\frac{\log n}{n}\right)$.
In contrast, in the Landau-rate regime ($\alpha=\beta$), the residual
term is bounded in magnitude by $\frac{2}{\sqrt{\small\mathsf{SNR}}}+\frac{\log\left(n+1\right)}{n}$.
In fact, the term $\mathcal{O}\left(\frac{\log n}{n}\right)$ arises
when using the entropy function to approximate the rate of binomial
coefficients, while an additional approximation loss $\mathcal{O}\left(\frac{1}{\small\mathsf{SNR}}\right)$
occurs when employing $\log\mathsf{SNR}$ to approximate $\log\left(1+\mathsf{SNR}\right)$.

\subsection{Achievability \label{sub:Achievability-Super-Landau-Minimax}}

In general, it is computationally intractable to find a deterministic
solution to approach the minimax limits by solving a numerical optimization
program. Fortunately, when $n$ and SNR are both large, simple random
sampling strategies suffice in approaching the minimax information
rate loss limits uniformly under all channel occupancy. The achievability
result is formally stated in the following theorem. 

\begin{theorem}\label{theorem-Minimax-Sampler-BWLimited-Achievability-Oversampling}Let
$\boldsymbol{M}\in\mathbb{R}^{m\times n}$ be a Gaussian matrix such
that $\boldsymbol{M}_{ij}$'s are independently drawn from $\mathcal{N}\left(0,1\right)$.
Then with probability exceeding $1-C\exp\left(-n\right)$, the following
holds:

(a) If $\beta<\alpha\text{ and }\alpha+\beta<1$, then 
\begin{equation}
\max_{\boldsymbol{s}\in{[n] \choose k}}L_{\boldsymbol{s}}^{\boldsymbol{M}}\leq\frac{W}{2}\left\{ \mathcal{H}(\beta)-\alpha\mathcal{H}\left(\frac{\beta}{\alpha}\right)+\Lambda+\frac{c_{1}\log n}{n^{1/3}}\right\} ;
\end{equation}

(b) If $\alpha<\beta$, then 
\begin{equation}
\max_{\boldsymbol{s}\in{[n] \choose k}}L_{\boldsymbol{s}}^{\boldsymbol{M}}\leq\frac{W}{2}\left\{ \mathcal{H}(\alpha)-\beta\mathcal{H}\left(\frac{\alpha}{\beta}\right)+\Lambda+\frac{c_{2}\log n}{n^{1/3}}\right\} ;
\end{equation}

(c) If $\beta=\alpha$ or if $\beta<\alpha\text{ and }\alpha+\beta\geq1$,
then
\begin{equation}
\max_{\boldsymbol{s}\in{[n] \choose k}}L_{\boldsymbol{s}}^{\boldsymbol{M}}\leq\frac{W}{2}\left\{ \mathcal{H}(\beta)-\alpha\mathcal{H}\left(\frac{\beta}{\alpha}\right)+\Lambda+\frac{c_{3}\mathsf{SNR}^{\frac{1}{3}}\log n}{n^{1/3}}\right\} .
\end{equation}

Here, $C,c_{1}\sim c_{3}>0$ are some universal constants independent
of $\mathrm{SNR}$ and $n$, and $\Lambda$ is given in (\ref{eq:Delta_alpha_beta}).\end{theorem}

Theorem \ref{theorem-Minimax-Sampler-BWLimited-Achievability-Oversampling}
indicates that Gaussian sampling approaches the minimax information
rate loss (which is about $\frac{1}{2}\mathcal{H}(\beta)-\frac{1}{2}\alpha\mathcal{H}\left(\frac{\beta}{\alpha}\right)$
per Hertz) to within a small gap. In fact, with exponentially high
probability, the sampled information rate loss is almost equivalent
to the  minimax limit uniformly across all states $\boldsymbol{s}\in{[n] \choose k}$,
as will be shown later.

\section{Equivalent Algebraic Problems\label{sec:Equivalent-Algebraic-Problems}}

Our main results in Section \ref{sec:Minimax-Capacity-Regret} can
be established by investigating equivalent algebraic problems. Observe
that the information rate loss can be expressed as
\begin{align}
 & L_{\boldsymbol{s}}^{\boldsymbol{Q}}=-C_{\boldsymbol{s}}^{\boldsymbol{Q}}+C_{\boldsymbol{s}}\nonumber \\
 & \text{ }=-\frac{W}{2n}\log\det\left(\boldsymbol{I}_{m}+\mathsf{SNR}\cdot\boldsymbol{Q}_{\boldsymbol{s}}^{\text{w}}\boldsymbol{Q}_{\boldsymbol{s}}^{\text{w}*}\right)\nonumber \\
 & \quad\quad+\frac{W\min\left\{ k,m\right\} }{2n}\left\{ \log\mathsf{SNR}+\log\left(1+\frac{1}{\small\mathsf{SNR}}\right)\right\} \\
 & \text{ }=\begin{cases}
\frac{W}{2}\left\{ -\frac{1}{n}\log\det\left(\frac{1}{\small\mathsf{SNR}}\boldsymbol{I}_{k}+\boldsymbol{Q}_{\boldsymbol{s}}^{\text{w}*}\boldsymbol{Q}_{\boldsymbol{s}}^{\text{w}}\right)+\Lambda\right\} , & \text{if }k\leq m\\
\frac{W}{2}\left\{ -\frac{1}{n}\log\det\left(\frac{1}{\small\mathsf{SNR}}\boldsymbol{I}_{m}+\boldsymbol{Q}_{\boldsymbol{s}}^{\text{w}}\boldsymbol{Q}_{\boldsymbol{s}}^{\text{w}*}\right)+\Lambda\right\} , & \text{if }k\geq m
\end{cases}\label{eq:GsQApproximationLandauSampling}
\end{align}
where $\Lambda=\mathcal{O}\left(\frac{1}{\small\mathsf{SNR}}\right)$
is defined in (\ref{eq:Delta_alpha_beta}). This identity makes $\log\det\left(\epsilon\boldsymbol{I}_{k}+\boldsymbol{Q}_{\boldsymbol{s}}^{\text{w}*}\boldsymbol{Q}_{\boldsymbol{s}}^{\text{w}}\right)$
a quantity of interest. In the sequel, we provide tight bounds on
this quantity, which in turn establish Theorems \ref{Theorem-Minimax-Det-Upper-Bound}-\ref{theorem-Minimax-Sampler-BWLimited-Achievability-Oversampling}.
The proofs of these results rely heavily on \emph{non-asymptotic}
(random) matrix theory. In particular, the proofs for the achievability
bounds are established based on measure concentration of log-determinant
functions, which will be provided at the end of this section.

\subsection{Upper Bound on Log Determinants}

Recall that $\boldsymbol{Q}^{\text{w}}\boldsymbol{Q}^{\text{w}*}=\boldsymbol{I}$,
and that $\boldsymbol{B}_{\boldsymbol{s}}$ represents the $m\times k$
submatrix of $\boldsymbol{B}$ with columns coming from the index
set $\boldsymbol{s}$. The following theorem investigates the properties
of $\log\det\left(\epsilon\boldsymbol{I}_{k}+\boldsymbol{B}_{\boldsymbol{s}}^{*}\boldsymbol{B}_{\boldsymbol{s}}\right)$
for any $m\times n$ matrix $\boldsymbol{B}$ that has orthonormal
rows. 

\begin{theorem}\label{theorem-Evaluate-Det-IplusBB}Consider any
$\epsilon>0$. Let $\boldsymbol{B}$ be any $m\times n$ matrix that
satisfies $\boldsymbol{B}\boldsymbol{B}^{*}=\boldsymbol{I}_{m}$.

(1) If $\beta\leq\alpha\leq1$, then
\begin{align}
 & \min_{\boldsymbol{s}\in{[n] \choose k}}\log\det\left(\epsilon\boldsymbol{I}_{k}+\boldsymbol{B}_{\boldsymbol{s}}^{*}\boldsymbol{B}_{\boldsymbol{s}}\right)\nonumber \\
 & \text{ }\leq\alpha\mathcal{H}\left(\frac{\beta}{\alpha}\right)-\mathcal{H}\left(\beta\right)+\frac{\log\left(n+1\right)}{n}+\nonumber \\
 & \small\text{ }\text{ }\min\left\{ \left(1+\beta\right)\log\left(1+\sqrt{\epsilon}\right),\left\lceil \frac{\beta}{\alpha-\beta}\right\rceil \log\left(1+\epsilon\right)\right\} .\label{eq:MaximinLogDetIplusBB}
\end{align}

(2) If $\alpha\leq\beta\leq1$, then
\begin{align}
 & \min_{\boldsymbol{s}\in{[n] \choose k}}\log\det\left(\epsilon\boldsymbol{I}_{m}+\boldsymbol{B}_{\boldsymbol{s}}\boldsymbol{B}_{\boldsymbol{s}}^{*}\right)\nonumber \\
 & \text{ }\leq\beta\mathcal{H}\left(\frac{\alpha}{\beta}\right)-\mathcal{H}\left(\alpha\right)+\frac{\log\left(n+1\right)}{n}+\nonumber \\
 & \small\quad\min\left\{ \left(1+\alpha\right)\log\left(1+\sqrt{\epsilon}\right),\left\lceil \frac{\left(1-\alpha\right)\alpha}{\beta-\alpha}\right\rceil \log\left(1+\epsilon\right)\right\} .\label{eq:MaximinLogDetIplusBB-sub}
\end{align}
\end{theorem}

Theorem \ref{theorem-Evaluate-Det-IplusBB} together with (\ref{eq:GsQApproximationLandauSampling})
suggests that
\begin{align*}
\frac{L}{W/2} & \geq\begin{cases}
\mathcal{H}\left(\beta\right)-\alpha\mathcal{H}\left(\frac{\beta}{\alpha}\right)-\mathcal{O}\left(\frac{1}{\small\mathsf{SNR}}\right)-\frac{\log\left(n+1\right)}{n},\\
\quad\quad\quad\quad\quad\quad\quad\quad\quad\quad\quad\quad\quad\quad\quad\quad\text{if }\alpha\geq\beta,\\
\mathcal{H}\left(\alpha\right)-\beta\mathcal{H}\left(\frac{\alpha}{\beta}\right)-\mathcal{O}\left(\frac{1}{\small\mathsf{SNR}}\right)-\frac{\log\left(n+1\right)}{n},\\
\quad\quad\quad\quad\quad\quad\quad\quad\quad\quad\quad\quad\quad\quad\quad\quad\text{if }\alpha<\beta,
\end{cases}
\end{align*}
which completes the proof of Theorem \ref{Theorem-Minimax-Det-Upper-Bound}.

One of the key ingredients in establishing Theorem \ref{theorem-Evaluate-Det-IplusBB}
is to demonstrate that the sum 
\begin{equation}
\sum_{\boldsymbol{s}\in{[n] \choose k}}\det\left(\epsilon\boldsymbol{I}+\boldsymbol{B}_{\boldsymbol{s}}^{*}\boldsymbol{B}_{\boldsymbol{s}}\right)
\end{equation}
is a \emph{constant} independent of the matrix $\boldsymbol{B}$,
as long as $\boldsymbol{B}$ has orthonormal rows. Consequently, in
order to maximize $\min_{\boldsymbol{s}}\det\left(\epsilon\boldsymbol{I}+\boldsymbol{B}_{\boldsymbol{s}}^{*}\boldsymbol{B}_{\boldsymbol{s}}\right)$,
one would wish to find a matrix $\boldsymbol{B}$ such that $\det\left(\epsilon\boldsymbol{I}+\boldsymbol{B}_{\boldsymbol{s}}^{*}\boldsymbol{B}_{\boldsymbol{s}}\right)$
are almost identical over all $\boldsymbol{s}$. When translated to
the language of channel capacity, this observation suggests that an
ideal minimax sampling method should be able to achieve (almost) equivalent
information rate loss uniformly over all states $\boldsymbol{s}$,
for which random sampling becomes a natural candidate due to sharp
concentration of measures.

\subsection{Achievability under Gaussian Ensembles}

When it comes to the achievability part, the major step is to quantify
$\log\det(\epsilon\boldsymbol{I}+(\boldsymbol{M}\boldsymbol{M}^{\top})^{-1}\boldsymbol{M}_{\boldsymbol{s}}\boldsymbol{M}_{\boldsymbol{s}}^{\top})$
for every $\boldsymbol{s}\in{[n] \choose k}$. Interestingly, this
quantity can be uniformly controlled due to the concentration of spectral
measure of random matrices \cite{GuionnetZeitouni2000}. This is stated
in the following theorem, which demonstrates the optimality of Gaussian
sampling mechanisms.

\begin{theorem}\label{theorem-Minimax-RMT-SuperLandau}Consider any
$\epsilon>0$. Let $\boldsymbol{M}\in\mathbb{R}^{m\times n}$ be an
i.i.d. random matrix satisfying $\boldsymbol{M}_{ij}\sim\mathcal{N}(0,1)$. 

(a) If $\beta<\alpha$ and $\alpha+\beta<1$, then with probability
at least $1-C\exp\left(-n\right)$,
\begin{align}
 & \min_{\boldsymbol{s}\in{[n] \choose k}}\frac{1}{n}\log\det\left(\epsilon\boldsymbol{I}_{k}+\boldsymbol{M}_{\boldsymbol{s}}^{\top}\left(\boldsymbol{M}\boldsymbol{M}^{\top}\right)^{-1}\boldsymbol{M}_{\boldsymbol{s}}\right)\nonumber \\
 & \quad\geq-\mathcal{H}(\beta)+\alpha\mathcal{H}\left(\frac{\beta}{\alpha}\right)+\frac{c_{1}\log n}{n^{1/3}}.
\end{align}

(b) If $\alpha<\beta\leq1$, then with probability at least $1-9\exp\left(-2n\right)$,
\begin{align}
 & \min_{\boldsymbol{s}\in{[n] \choose k}}\frac{1}{n}\log\det\left(\epsilon\boldsymbol{I}_{m}+(\boldsymbol{M}\boldsymbol{M}^{\top})^{-\frac{1}{2}}\boldsymbol{M}_{\boldsymbol{s}}\boldsymbol{M}_{\boldsymbol{s}}^{\top}(\boldsymbol{M}\boldsymbol{M}^{\top})^{-\frac{1}{2}}\right)\nonumber \\
 & \quad\geq\beta\mathcal{H}\left(\frac{\alpha}{\beta}\right)-\mathcal{H}\left(\alpha\right)-\frac{c_{2}\log n}{n^{1/3}}.
\end{align}

(c) If $\beta=\alpha$ or if $\beta<\alpha$ and $\alpha+\beta\geq1$,
then with probability exceeding $1-9\exp\left(-2n\right)$, 
\begin{align}
 & \min_{\boldsymbol{s}\in{[n] \choose k}}\frac{1}{n}\log\det\left(\epsilon\boldsymbol{I}_{k}+\boldsymbol{M}_{\boldsymbol{s}}^{\top}\left(\boldsymbol{M}\boldsymbol{M}^{\top}\right)^{-1}\boldsymbol{M}_{\boldsymbol{s}}\right)\nonumber \\
 & \quad\geq\text{ }\alpha\mathcal{H}\left(\frac{\beta}{\alpha}\right)-\mathcal{H}\left(\beta\right)-\frac{c_{3}\log n}{\left(\epsilon n\right)^{1/3}}.
\end{align}

Here, $c_{1},c_{2},c_{3},C>0$ are some universal constants independent
of $n$ and $\epsilon$.

\end{theorem}

Putting Theorem \ref{theorem-Minimax-RMT-SuperLandau} and Equation
(\ref{eq:GsQApproximationLandauSampling}) together implies that $\forall\boldsymbol{s}\in{[n] \choose k}$,
\[
\frac{L_{\boldsymbol{s}}^{\boldsymbol{M}}}{W/2}\leq\begin{cases}
\mathcal{H}(\beta)-\alpha\mathcal{H}\left(\frac{\beta}{\alpha}\right)+\Lambda+\mathcal{O}\left(\frac{\log n}{n^{\frac{1}{3}}}\right),\quad\\
\quad\quad\quad\quad\quad\quad\text{if }\beta<\alpha\text{ and }\alpha+\beta<1\\
\mathcal{H}\left(\alpha\right)-\beta\mathcal{H}\left(\frac{\alpha}{\beta}\right)+\Lambda+\mathcal{O}\left(\frac{\log n}{n^{\frac{1}{3}}}\right),\\
\quad\quad\quad\quad\quad\quad\text{if }\beta>\alpha\\
\mathcal{H}\left(\beta\right)-\alpha\mathcal{H}\left(\frac{\beta}{\alpha}\right)+\Lambda+\mathcal{O}\left(\frac{\mathsf{SNR}^{\frac{1}{3}}\log n}{n^{\frac{1}{3}}}\right),\\
\quad\quad\quad\quad\quad\quad\text{if }\beta=\alpha\text{ or if }\beta<\alpha\text{ and }\alpha+\beta\geq1
\end{cases}
\]
with exponentially high probability, which establishes Theorem \ref{theorem-Minimax-Sampler-BWLimited-Achievability-Oversampling}.
The above achievability bounds are established via the concentration
of spectral measure of large random matrices. 

\subsection{Measure Concentration of Log-Determinant functions for Random Matrices\label{sub:Concentration-of-Log-Determinant}}

As mentioned above, the key machinery in establishing the achievability
bounds is to evaluate certain log-determinant functions. In fact,
many limiting results for i.i.d. Gaussian ensembles have been derived
when studying MIMO fading channels (e.g. \cite{Tel1999,lozano2002capacity,chuah2002capacity,TulinoVerdu2004}),
which focus on the first-order limits instead of the convergence rate.
Furthermore, the second-order asymptotics and the large deviation
for mutual information have also been studied (e.g. \cite{hachem2008new,kazakopoulos2011living})
in the asymptotic regime of large $n$. Most of these results focus
on a special case of the log-determinant function (i.e. $\log\det(\epsilon\boldsymbol{I}+\frac{1}{n}\boldsymbol{M}\boldsymbol{M}^{\top})$)
and suppose that $n$ scales independent of $\epsilon$. On the other
hand, the concentration of some log-determinant functions has been
studied in the random matrix literature as a key step in establishing
universal laws for linear spectral statistics (e.g. \cite[Proposition 48]{TaoVu2012Nonhermitian}).
However, these bounds are only shown to hold with overwhelming probability
(i.e. with probability $1-e^{-\omega\left(\log n\right)}$), which
are not sharp enough for our purpose. As a result, we provide sharper
measure concentration results of log-determinants in this subsection.

One important class of log-determinant functions takes the form of
$\frac{1}{n}\log\det\left(\frac{1}{n}\boldsymbol{A}\boldsymbol{A}^{\top}\right)$.
The concentration of such functions for i.i.d. rectangular Gaussian
matrices is characterized in the following lemmas.

\begin{lem}\label{lem:GaussianLocalSpectrum}Suppose that $\boldsymbol{A}\in\mathbb{R}^{m\times n}$
is a random matrix whose entries are independent standard Gaussian
random variables. Assume that $0<\alpha<1$. Then for any $\delta>0$
and any $\tau>0$,
\begin{equation}
\frac{\mathrm{card}\left\{ i\mid\lambda_{i}\left(\frac{1}{n}\boldsymbol{A}\boldsymbol{A}^{\top}\right)<\delta\right\} }{n}<\frac{\alpha}{1-\alpha-\frac{1}{n}}\delta+\frac{4\sqrt{\alpha\tau}}{\sqrt{n\delta}}\label{eq:SmallEvalueCardinalityBound-1}
\end{equation}
holds with probability exceeding $1-2\exp\left(-\tau n\right)$. \end{lem}

\begin{proof}See Appendix \ref{sec:Proof-of-Lemma-Gaussian-Local-Spectral}.\end{proof}

\begin{lem}\label{lem:GaussianLogDet_LB}Suppose that $\boldsymbol{A}\in\mathbb{R}^{m\times n}$
is a random matrix with independent entries satisfying $\boldsymbol{A}_{ij}\sim\mathcal{N}(0,1)$.
Assume that $0<\alpha<1$. 

(1) For any $\tau>0$ and any $n>\max\left\{ \frac{2}{1-\sqrt{\alpha}},\frac{2}{\tau},7\right\} $,
\begin{align}
\frac{1}{n}\log\det\left(\frac{1}{n}\boldsymbol{A}\boldsymbol{A}^{\top}\right) & \leq\left(1-\alpha\right)\log\frac{1}{1-\alpha}-\alpha+\frac{2\log n}{n}\nonumber \\
 & \quad\quad+\frac{5\sqrt{\alpha}}{\left(1-\sqrt{\alpha}-\frac{2}{n}\right)}\frac{\tau}{\sqrt{n}}\label{eq:logdetAA_UB_Gaussian}
\end{align}
with probability exceeding $1-2\exp\left(-2\tau^{2}n\right)$.

(2) For any $n>\max\left\{ \frac{6.414}{1-\alpha}\cdot e^{\frac{\tau^{2}}{1-\alpha}},\left(\frac{\alpha}{1-\alpha-\frac{1}{n}}+4\sqrt{\alpha}\tau\right)^{3}\right\} $
and any $\tau>0$,
\begin{align}
\frac{1}{n}\log\det\left(\frac{1}{n}\boldsymbol{A}\boldsymbol{A}^{\top}\right) & \geq\left(1-\alpha\right)\log\frac{1}{1-\alpha}-\alpha\nonumber \\
 & \quad\quad-\frac{\left(\frac{2}{1-\alpha-\frac{1}{n}}+10\tau\right)\log n}{n^{1/3}}\label{eq:Gaussian_LB_logdetAA}
\end{align}
with probability exceeding $1-7\exp(-\tau^{2}n)$.\end{lem}

\begin{proof}See Appendix \ref{sec:Proof-of-Lemma-GaussianLogDet_LB}.\end{proof}

The last log-determinant function considered here takes the form of
$\log\det\left(\epsilon\boldsymbol{I}+\boldsymbol{A}^{\top}\boldsymbol{B}^{-1}\boldsymbol{A}\right)$
for some independent random matrices $\boldsymbol{A}$ and $\boldsymbol{B}$,
as stated in the following lemma.

\begin{lem}\label{lemma-Expected-Log-Determinant-LambdaStatistics}Suppose
that $\beta<\alpha$ and $\alpha+\beta\leq1$. Let $\boldsymbol{A}=\mathbb{R}^{m\times k}$
be a random matrix whose entries are independent standard Gaussian
random variables, and let $\boldsymbol{B}\sim\mathcal{W}_{m}\left(n-k,\boldsymbol{I}_{m}\right)$
be independent of $\boldsymbol{A}$. Then for any $\tau>0$,
\begin{align}
 & \frac{1}{n}\log\det\left(\epsilon\boldsymbol{I}_{k}+\boldsymbol{A}^{\top}\boldsymbol{B}^{-1}\boldsymbol{A}\right)\nonumber \\
 & \quad\geq-\left(\alpha-\beta\right)\log\left(\alpha-\beta\right)+\alpha\log\alpha-\beta\log\left(1-\alpha\right)\nonumber \\
 & \quad\quad+\left(1-\alpha-\beta\right)\log\left(1-\frac{\beta}{1-\alpha}\right)-\frac{\left(c_{8}+c_{9}\tau\right)\log n}{n^{1/3}}
\end{align}
with probability exceeding $1-9\exp\left(-\tau^{2}n\right)$, where
$c_{8},c_{9}>0$ are some universal constants\footnote{More precisely, by setting $\zeta:=\max\left\{ \frac{\beta}{\alpha},\frac{\alpha}{1-\beta}\right\} $,
one can take $c_{8}=\frac{3}{1-\zeta}$ and $c_{9}=\frac{8\left(2-\sqrt{\zeta}\right)}{1-\sqrt{\zeta}}$
for sufficiently large $n$.} independent of $n$ and $\epsilon$. \end{lem}

\begin{proof}See Appendix \ref{sec:Proof-of-lemma-Expected-Log-Determinant-LambdaStatistics}.\end{proof}

As demonstrated in (\ref{eq:GsQApproximationLandauSampling}), the
information rate loss is captured by some logarithmic function. The
preceding concentration of measure results will prove useful in determining
such a information rate loss metric.

\section{Discussion\label{sec:Discussion}}

\subsection{Implications of Main Results \label{sub:Implications-Minimax}}

In this subsection, we summarize several key insights from the main
theorems. 

\begin{figure}
\begin{centering}
\includegraphics[scale=0.45]{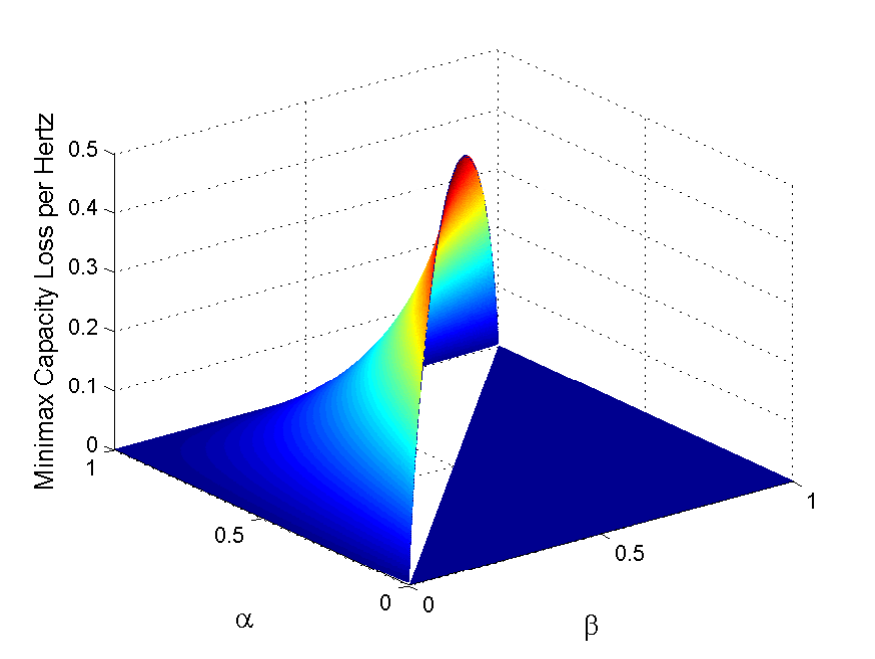}
\par\end{centering}

\begin{centering}
\includegraphics[scale=0.45]{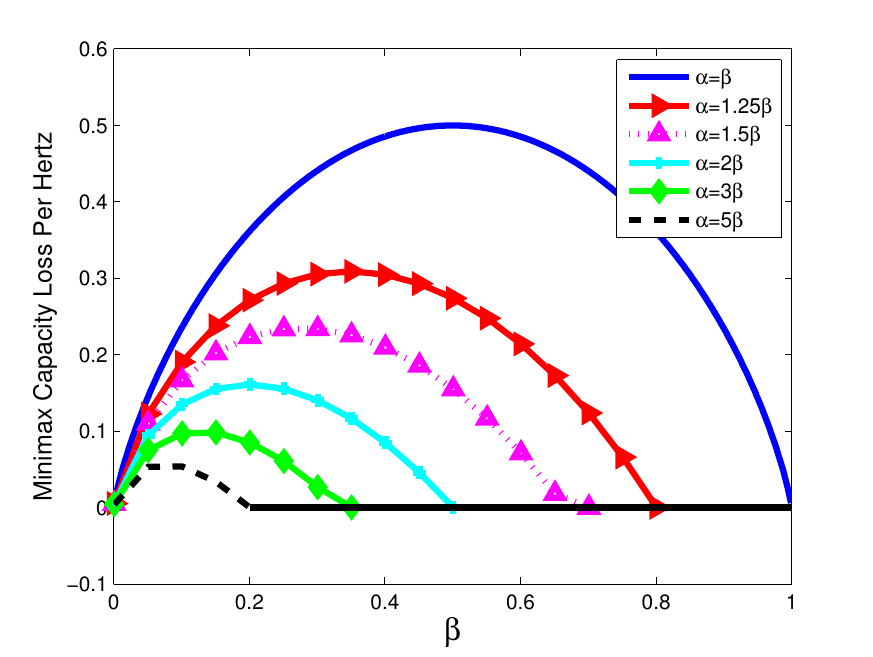}
\par\end{centering}

\caption{\label{fig:SuperLandauBinaryEntropyFunction}The minimax loss per
Hertz (without residual terms) vs. the sparsity factor $\beta$ and
the undersampling factor $\alpha$. }
\end{figure}

\begin{itemize}[listparindent =1em]\itemsep0.5em \item[(1)] For the
whole sub-Nyquist regime, the minimax information rate loss is captured
by several binary entropy functions. When the number of subbands and
the SNR are sufficiently large and $\frac{\small\mathsf{SNR}}{n}=o\left(1\right)$,
the minimax limits depend almost only on the undersampling factor
and the sparsity factor rather than $\left(n,k,m\right)$, which are
plotted in Fig. \ref{fig:SuperLandauBinaryEntropyFunction}. It can
be observed from the plot that when sampling above the Landau rate
(but below the Nyquist rate), increasing the $\alpha/\beta$ ratio
improves the capacity gap, and shrinks the locus. In contrast, when
$\alpha<\beta$, increasing $\alpha/\beta$ results in a worse capacity
gap. This implies that in the sub-Landau regime, the \emph{relative
information rate loss} is easier to control when $\alpha/\beta$ decreases,
although the achievable channel capacity also shrinks. In fact, the
information rate loss is the largest under Landau-rate sampling, as
illustrated in Fig. \ref{fig:SuperLandauBinaryEntropyFunction}. Since
the capacity under channel-optimized sampling scales as $\Theta\left(W\log\mathsf{SNR}\right)$,
our results indicate that the ratio of the minimax information rate
loss to the maximum capacity vanishes at a rate $\Theta\left(1/\log\mathsf{SNR}\right)$. 

\item[(2)] Note that under Landau-rate sampling (i.e. $\alpha=\beta$),
the minimax loss is $\frac{1}{2}\mathcal{H}\left(\beta\right)$ (or
$\frac{1}{2}\mathcal{H}\left(\alpha\right)$) modulo some residual
terms. As a result, if we fix the channel sparsity factor and increase
the sampling rate above the Landau rate, then the capacity benefit
per unit bandwidth is captured by the term $\frac{1}{2}\alpha\mathcal{H}\left(\beta/\alpha\right)$.
On the other hand, if we fix the sampling rate but increase the channel
occupancy, then the capacity gain per Hertz one can harvest amounts
to $\frac{1}{2}\beta\mathcal{H}\left(\alpha/\beta\right)$. For either
case, if $\alpha\rightarrow1$, the information rate loss per Hertz
reduces to
\[
\frac{1}{2}\mathcal{H}\left(\beta\right)-\frac{1}{2}\alpha\mathcal{H}\left(\frac{\beta}{\alpha}\right)=0,
\]
meaning that there is effectively no information rate loss under Nyquist-rate
sampling. This agrees with the fact that Nyquist-rate sampling is
information preserving.

\item[(3)] The information rate loss incurred by Gaussian random
sampling meets the  minimax limit for Landau-rate sampling uniformly
across all states $\boldsymbol{s}$, which reveals that with exponentially
high probability, random sampling is optimal in terms of universal
sampling design. This arises since the capacity achievable by random
sampling exhibits very sharp measure concentration. \end{itemize}

\subsection{Extension \label{sub:Extension}}

\begin{itemize}[listparindent =1em]\itemsep0.5em \item \textbf{Universality
of Random Sampling Schemes} \textbf{Beyond Gaussian Sampling}. While
the main theorems provided in the present paper focus on Gaussian
sampling, we remark that a much broader class of random sampling strategies
are also minimax-optimal. This subsumes the class of sampling coefficient
matrices $\boldsymbol{M}$ such that its entries are independent sub-Gaussian
random variables with matching moments up to the second order. The\emph{
universality} phenomenon that arises in large random matrices (e.g.
\cite{TaoVu2012Covariance}) suggests that the minimaxity of random
sampling matrices does not depend on the particular distribution of
the coefficients, although they might affect the convergence rate
to some degree. Interested readers are referred to \cite{chen2014Thesis}
for derivation of these results and associated insights.

\item \textbf{Beyond Frequency-Flat Channels and Uniform Power Allocation}.
The present paper concentrates on the frequency-flat channel models
for simplicity of presentation. Note, however, that the main results
derived herein can be readily extended to more general frequency-varying
channels. Specifically, suppose that the information rate loss metric
is defined as the gap of achievable rates under universal sampling
relative to channel-optimized sampling (both employing optimal power
allocation). Then as long as the peak-to-average SNR
\[
\frac{\sup_{f\in\left[0,W\right]}\left|H\left(f\right)\right|^{2}/\mathcal{S}_{\eta}\left(f\right)}{\frac{1}{W}\int_{0}^{W}\left|H\left(f\right)\right|^{2}/\mathcal{S}_{\eta}\left(f\right)}
\]
is bounded, all results presented in this paper still hold, except
for some additional gap on the order of $\mathcal{O}\left(\frac{1}{\small\mathsf{SNR}_{\min}}\right)$,
where
\[
\mathsf{SNR}_{\min}:=\inf_{f\in\left[0,W\right]}\frac{P}{\min\left\{ \alpha,\beta\right\} W}\frac{\left|H\left(f\right)\right|^{2}}{\mathcal{S}_{\eta}\left(f\right)}.
\]
A proof of this result and derivation of the optimal power allocation
is provided in \cite{chen2014Thesis}.

\end{itemize}

\section{Conclusions\label{sec:Conclusion-Minimax}}

We have investigated minimax universal sampling design from a capacity
perspective. In order to characterize the loss due to universal sub-Nyquist
sampling design, we introduced the notion of sampled information rate
loss relative to the capacity under channel-optimized sampling, and
characterize overall robustness of the sampling design through the
minimax information rate loss metric. Specifically, we have determined
the minimax limit on the sampled information rate loss achievable
by a class of channel-blind periodic sampling systems. This minimax
limit turns out to be a constant that depends solely on the band sparsity
factor and undersampling factor, modulo some residual term that vanishes
as the SNR and the number of subbands grow. Our results demonstrate
that with exponentially high probability, Gaussian random sampling
is minimax-optimal in terms of a channel-blind sampler design. 

It remains to study how to extend this framework to situations beyond
compound multiband channels. Our notion of sampled information rate
loss might be useful in studying the robustness for these scenarios.
Our framework and results may also be appropriate for other channels
with state where sparsity exists in other transform domains. In addition,
when it comes to multiple access channels or random access channels
\cite{ElGamalKim2012}, it would be interesting to see how to design
a channel-blind sampler that is robust for the entire capacity region.

\section*{Acknowledgments}

The authors would like to thank Prof. Young-Han Kim and the anonymous
reviewer for extensive and constructive comments that greatly improved
the presentation and positioning of the paper. Y. Chen would like
to thank Dr. Ping Xia for discussion of LTE.

\appendices

\section{Proof of Theorem \ref{theorem-Evaluate-Det-IplusBB}\label{sec:Proof-of-Theorem-Det-I-plus-BB}}

Before proving the results, we first state two facts. Consider any
$m\times m$ matrix $\boldsymbol{A}$, and list the eigenvalues of
$\boldsymbol{A}$ as $\lambda_{1},\cdots,\lambda_{m}$. Define the
characteristic polynomial of $\boldsymbol{A}$ as
\begin{align}
p_{\boldsymbol{A}}(t) & =\det\left(t\boldsymbol{I}-\boldsymbol{A}\right)\nonumber \\
 & =t^{m}-S_{1}t^{m-1}+\cdots+(-1)^{m}S_{m},
\end{align}
where $S_{l}$ is the $l$th elementary symmetric function of $\lambda_{1},\cdots,\lambda_{m}$
defined as follows: 
\begin{equation}
S_{l}:=\sum_{1\leq i_{1}<\cdots<i_{l}\leq m}\prod_{j=1}^{l}\lambda_{i_{j}}.
\end{equation}
We also define $E_{l}(\boldsymbol{A})$ as the sum of determinants
of all $l$-by-$l$ principal minors of $\boldsymbol{A}$. According
to \cite[Theorem 1.2.12]{HornJohnson}, $S_{l}=E_{l}(\boldsymbol{A})$
holds for all $1\leq l\leq m$, indicating that
\begin{equation}
\det\left(t\boldsymbol{I}+\boldsymbol{A}\right)=t^{m}+E_{1}(\boldsymbol{A})t^{m-1}+\cdots+E_{m}(\boldsymbol{A}).\label{eq:DetIplusA_via_PrincipalMinor}
\end{equation}

Another fact we will rely on is the entropy formula of binomial coefficients
\cite[Example 11.1.3]{cover2012elements}, that is, for every $0\leq k\leq n$,
\begin{equation}
\mathcal{H}\left(\frac{k}{n}\right)-\frac{\log(n+1)}{n}\leq\frac{1}{n}\log{n \choose k}\leq\mathcal{H}\left(\frac{k}{n}\right),\label{eq:EntropyApproximation}
\end{equation}
where $\mathcal{H}(x):=x\log\frac{1}{x}+(1-x)\log\frac{1}{1-x}$ stands
for the entropy function.

Now we are in position to prove the theorem. For any $m\times n$
matrix $\boldsymbol{B}$ obeying $\boldsymbol{B}\boldsymbol{B}^{*}=\boldsymbol{I}_{m}$,
using the identity (\ref{eq:DetIplusA_via_PrincipalMinor}) we get
\begin{align}
 & \sum_{\boldsymbol{s}\in{[n] \choose k}}\det\left(\epsilon\boldsymbol{I}+\boldsymbol{B}_{\boldsymbol{s}}\boldsymbol{B}_{\boldsymbol{s}}^{*}\right)\nonumber \\
 & \quad=\sum_{\boldsymbol{s}\in{[n] \choose k}}\left\{ \epsilon^{m}+\sum_{l=1}^{m}\epsilon^{m-l}E_{l}\left(\boldsymbol{B}_{\boldsymbol{s}}\boldsymbol{B}_{\boldsymbol{s}}^{*}\right)\right\} ,\label{eq:DecompositionFirst}\\
 & \quad=\epsilon^{m}{n \choose k}+\sum_{l=1}^{m}\epsilon^{m-l}\sum_{\boldsymbol{s}\in{[n] \choose k}}E_{l}\left(\boldsymbol{B}_{\boldsymbol{s}}\boldsymbol{B}_{\boldsymbol{s}}^{*}\right).\label{eq:DecompositionPrincipalMinor}
\end{align}
Consider an index set $\boldsymbol{r}\in{[m] \choose l}$ with $l\leq\min\left\{ k,m\right\} $,
and denote by $\left(\boldsymbol{B}_{\boldsymbol{s}}\boldsymbol{B}_{\boldsymbol{s}}^{*}\right)_{\boldsymbol{r}}$
the submatrix of $\boldsymbol{B}_{\boldsymbol{s}}\boldsymbol{B}_{\boldsymbol{s}}^{*}$
with rows and columns coming from the index set $\boldsymbol{r}$.
It can be verified that
\[
\det\left(\left(\boldsymbol{B}_{\boldsymbol{s}}\boldsymbol{B}_{\boldsymbol{s}}^{*}\right)_{\boldsymbol{r}}\right)=\det\left(\boldsymbol{B}_{\boldsymbol{r},\boldsymbol{s}}\boldsymbol{B}_{\boldsymbol{r},\boldsymbol{s}}^{*}\right)=\sum_{\tilde{\boldsymbol{r}}\in{\boldsymbol{s} \choose l}}\det\left(\boldsymbol{B}_{\boldsymbol{r},\tilde{\boldsymbol{r}}}\boldsymbol{B}_{\boldsymbol{r},\tilde{\boldsymbol{r}}}^{*}\right),
\]
where the last equality is a consequence of the Cauchy-Binet formula
(e.g. \cite{Tao2012RMT}). Some algebraic manipulation yields that
for any $l\leq\min\left\{ k,m\right\} $,
\begin{align}
 & \sum_{\boldsymbol{s}\in{[n] \choose k}}E_{l}\left(\boldsymbol{B}_{\boldsymbol{s}}\boldsymbol{B}_{\boldsymbol{s}}^{*}\right)=\sum_{\boldsymbol{s}\in{[n] \choose k}}\sum_{\boldsymbol{r}\in{[m] \choose l}}\det\left(\left(\boldsymbol{B}_{\boldsymbol{s}}\boldsymbol{B}_{\boldsymbol{s}}^{*}\right)_{\boldsymbol{r}}\right)\nonumber \\
 & \quad=\sum_{\boldsymbol{s}\in{[n] \choose k}}\sum_{\boldsymbol{r}\in{[m] \choose l}}\sum_{\tilde{\boldsymbol{r}}\in{\boldsymbol{s} \choose l}}\det\left(\boldsymbol{B}_{\boldsymbol{r},\tilde{\boldsymbol{r}}}\boldsymbol{B}_{\boldsymbol{r},\tilde{\boldsymbol{r}}}^{*}\right)\nonumber \\
 & \quad=\sum_{\boldsymbol{r}\in{[m] \choose l}}\sum_{\tilde{\boldsymbol{r}}\in{[n] \choose l}}\sum_{\boldsymbol{s}:\tilde{\boldsymbol{r}}\subseteq\boldsymbol{s}}\det\left(\boldsymbol{B}_{\boldsymbol{r},\tilde{\boldsymbol{r}}}\boldsymbol{B}_{\boldsymbol{r},\tilde{\boldsymbol{r}}}^{*}\right)\nonumber \\
 & \quad\overset{(a)}{=}\sum_{\boldsymbol{r}\in{[m] \choose l}}\sum_{\tilde{\boldsymbol{r}}\in{[n] \choose l}}{n-l \choose k-l}\det\left(\boldsymbol{B}_{\boldsymbol{r},\tilde{\boldsymbol{r}}}\boldsymbol{B}_{\boldsymbol{r},\tilde{\boldsymbol{r}}}^{*}\right)\nonumber \\
 & \quad\overset{(b)}{=}\sum_{\boldsymbol{r}\in{[m] \choose l}}{n-l \choose k-l}={n-l \choose k-l}{m \choose l},\label{eq:lthOrderPrincipalMinor}
\end{align}
where (a) follows since the number of $k$-combinations (out of $[n]$)
containing $\tilde{\boldsymbol{r}}$ (an $l$-combination) is ${n-l \choose k-l}$,
and $(b)$ arises from the Cauchy-Binet formula together with $\boldsymbol{B}\boldsymbol{B}^{*}=\boldsymbol{I}_{m}$,
i.e.
\[
\sum_{\tilde{\boldsymbol{r}}\in{[n] \choose l}}\det\left(\boldsymbol{B}_{\boldsymbol{r},\tilde{\boldsymbol{r}}}\boldsymbol{B}_{\boldsymbol{r},\tilde{\boldsymbol{r}}}^{*}\right)=\det\left(\boldsymbol{B}_{\boldsymbol{r},[n]}\boldsymbol{B}_{\boldsymbol{r},[n]}^{*}\right)=\det\left(\boldsymbol{I}_{l}\right)=1.
\]

(1) Suppose that $k\leq m\leq n$. The identity (\ref{eq:DecompositionPrincipalMinor})
reduces to
\begin{align}
 & \sum_{\boldsymbol{s}\in{[n] \choose k}}\det\left(\epsilon\boldsymbol{I}+\boldsymbol{B}_{\boldsymbol{s}}\boldsymbol{B}_{\boldsymbol{s}}^{*}\right)\nonumber \\
 & \quad=\epsilon^{m}{n \choose k}+\sum_{l=1}^{k}\epsilon^{m-l}\sum_{\boldsymbol{s}\in{[n] \choose k}}E_{l}\left(\boldsymbol{B}_{\boldsymbol{s}}\boldsymbol{B}_{\boldsymbol{s}}^{*}\right),\label{eq:DecompositionPrincipalMinor-1}
\end{align}
since any $l$th order ($l>k$) minor of $\boldsymbol{B}_{\boldsymbol{s}}\boldsymbol{B}_{\boldsymbol{s}}^{*}$
is rank deficient, i.e. $E_{l}\left(\boldsymbol{B}_{\boldsymbol{s}}\boldsymbol{B}_{\boldsymbol{s}}^{*}\right)=0$.
Substituting (\ref{eq:lthOrderPrincipalMinor}) into (\ref{eq:DecompositionPrincipalMinor-1})
yields 
\begin{align}
\sum_{\boldsymbol{s}\in{[n] \choose k}}\det\left(\epsilon\boldsymbol{I}+\boldsymbol{B}_{\boldsymbol{s}}\boldsymbol{B}_{\boldsymbol{s}}^{*}\right) & =\sum_{l=0}^{k}{n-l \choose k-l}{m \choose l}\epsilon^{m-l},
\end{align}
which further gives
\begin{align}
 & {n \choose k}\min_{\boldsymbol{s}\in{[n] \choose k}}\det\left(\epsilon\boldsymbol{I}+\boldsymbol{B}_{\boldsymbol{s}}^{*}\boldsymbol{B}_{\boldsymbol{s}}\right)\leq\sum_{\boldsymbol{s}\in{[n] \choose k}}\det\left(\epsilon\boldsymbol{I}+\boldsymbol{B}_{\boldsymbol{s}}^{*}\boldsymbol{B}_{\boldsymbol{s}}\right)\nonumber \\
 & \quad=\sum_{\boldsymbol{s}\in{[n] \choose k}}\epsilon^{k-m}\det\left(\epsilon\boldsymbol{I}+\boldsymbol{B}_{\boldsymbol{s}}\boldsymbol{B}_{\boldsymbol{s}}^{*}\right)\\
 & \quad=\sum_{l=0}^{k}{n-l \choose k-l}{m \choose l}\epsilon^{k-l}.\label{eq:DetBB-expression}
\end{align}

The above expression allows us to derive an upper bound as
\begin{align}
 & {n \choose k}\min_{\boldsymbol{s}\in{[n] \choose k}}\det\left(\epsilon\boldsymbol{I}+\boldsymbol{B}_{\boldsymbol{s}}^{*}\boldsymbol{B}_{\boldsymbol{s}}\right)\nonumber \\
 & \quad\leq\sum_{l=0}^{k}{n \choose k-l}{m \choose l}\epsilon^{k-l}=\sum_{l=0}^{k}{n \choose l}{m \choose m-k+l}\epsilon^{l}\nonumber \\
 & \quad={m \choose k}\sum_{l=0}^{k}{n \choose l}\frac{{m \choose m-k+l}}{{m \choose k}}\epsilon^{l}.\label{eq:det3}
\end{align}
Since the term ${m \choose m-k+l}/{m \choose k}$ can be bounded above
by
\begin{align*}
\frac{{m \choose m-k+l}}{{m \choose k}} & =\frac{\frac{m!}{\left(m-k+l\right)!\left(k-l\right)!}}{\frac{m!}{k!\left(m-k\right)!}}=\frac{k!}{\left(k-l\right)!\frac{\left(m-k+l\right)!}{\left(m-k\right)!}}\\
 & =\frac{{k \choose l}}{{m-k+l \choose l}}\leq{k \choose l},
\end{align*}
plugging this inequality into (\ref{eq:det3}) we obtain
\begin{align}
 & {n \choose k}\min_{\boldsymbol{s}\in{[n] \choose k}}\det\left(\epsilon\boldsymbol{I}+\boldsymbol{B}_{\boldsymbol{s}}^{*}\boldsymbol{B}_{\boldsymbol{s}}\right)\nonumber \\
 & \quad\leq{m \choose k}\sum_{l=0}^{k}{n \choose l}{k \choose l}\epsilon^{l}\leq{m \choose k}\sum_{l=0}^{k}{n+k \choose 2l}\left(\sqrt{\epsilon}\right)^{2l}\nonumber \\
 & \quad\leq{m \choose k}\left(1+\sqrt{\epsilon}\right)^{n+k},\label{eq:SumDetEpsilonBsBsUB}
\end{align}
where the last equality follows from the binomial theorem. 

In addition, the identity (\ref{eq:DetBB-expression}) can also be
written as
\begin{align}
\sum_{\boldsymbol{s}\in{[n] \choose k}}\det\left(\epsilon\boldsymbol{I}+\boldsymbol{B}_{\boldsymbol{s}}^{*}\boldsymbol{B}_{\boldsymbol{s}}\right) & =\sum_{l=0}^{k}{n-k+l \choose l}{m \choose k-l}\epsilon^{l}.\label{eq:det2}
\end{align}
For any integer $K\geq n$, one can verify that $\forall0\leq l\leq k$,
\begin{align}
\frac{{n-k+l \choose l}}{{K \choose l}}\frac{{m \choose k-l}}{{m \choose k}} & =\frac{\prod_{i=1}^{l}\left(n-k+i\right)}{\prod_{i=0}^{l-1}\left(K-i\right)}\cdot\frac{\prod_{i=1}^{l}\left(k-l+i\right)}{\prod_{i=0}^{l-1}\left(m-k+l-i\right)}\nonumber \\
 & \leq\frac{\left(n-k+l\right)^{l}}{\left(K-l+1\right)^{l}}\cdot\frac{k^{l}}{\left(m-k+1\right)^{l}}\nonumber \\
 & \leq\left(\frac{nk}{K\left(m-k\right)}\right)^{l}.
\end{align}
Consequently, if $K\geq\frac{k}{m-k}n=\frac{\beta}{\alpha-\beta}n$,
then
\begin{equation}
{n-k+l \choose l}{m \choose k-l}\leq{K \choose l}{m \choose k},
\end{equation}
which combined with (\ref{eq:det2}) reveals that
\begin{align}
 & {n \choose k}\min_{\boldsymbol{s}\in{[n] \choose k}}\det\left(\epsilon\boldsymbol{I}+\boldsymbol{B}_{\boldsymbol{s}}^{*}\boldsymbol{B}_{\boldsymbol{s}}\right)\nonumber \\
 & \quad\leq{m \choose k}\sum_{l=0}^{k}{K \choose l}\epsilon^{l}\leq{m \choose k}\left(1+\epsilon\right)^{K}.\label{eq:SumDetEpsilonBsBsUB-2}
\end{align}

Set $K=\left\lceil \frac{\beta}{\alpha-\beta}\right\rceil n$. Putting
the preceding bounds (\ref{eq:SumDetEpsilonBsBsUB}) and (\ref{eq:SumDetEpsilonBsBsUB-2})
together suggests that
\begin{align*}
 & \min_{\boldsymbol{s}\in{[n] \choose k}}\frac{1}{n}\log\det\left(\epsilon\boldsymbol{I}+\boldsymbol{B}_{\boldsymbol{s}}^{*}\boldsymbol{B}_{\boldsymbol{s}}\right)\\
 & \text{ }\leq\frac{1}{n}\log{m \choose k}-\frac{1}{n}\log{n \choose k}\\
 & \quad\quad\quad+\underset{:=\varrho_{1}\left(\epsilon\right)}{\underbrace{\min\left\{ \frac{n+k}{n}\log\left(1+\sqrt{\epsilon}\right),\frac{K}{n}\log\left(1+\epsilon\right)\right\} }}.
\end{align*}
The entropy formula (\ref{eq:EntropyApproximation}) then allows us
to simplify:
\begin{align}
 & \min_{\boldsymbol{s}\in{[n] \choose k}}\frac{1}{n}\log\det\left(\epsilon\boldsymbol{I}+\boldsymbol{B}_{\boldsymbol{s}}^{*}\boldsymbol{B}_{\boldsymbol{s}}\right)\nonumber \\
 & \quad\leq\frac{m}{n}\mathcal{H}\left(\frac{k}{m}\right)-\mathcal{H}\left(\frac{k}{n}\right)+\frac{\log\left(n+1\right)}{n}+\varrho_{1}\left(\epsilon\right)\\
 & \quad=\alpha\mathcal{H}\left(\frac{\beta}{\alpha}\right)-\mathcal{H}\left(\beta\right)+\varrho_{1}\left(\epsilon\right)+\frac{\log\left(n+1\right)}{n}\label{eq:UBLogDetEpsilonBsBs_Entropy}
\end{align}
as claimed.

(2) When $m\leq k$, the identity (\ref{eq:DecompositionPrincipalMinor})
combined with (\ref{eq:lthOrderPrincipalMinor}) leads to
\begin{align}
 & {n \choose k}\min_{\boldsymbol{s}\in{[n] \choose k}}\det\left(\epsilon\boldsymbol{I}+\boldsymbol{B}_{\boldsymbol{s}}\boldsymbol{B}_{\boldsymbol{s}}^{*}\right)\leq\sum_{\boldsymbol{s}\in{[n] \choose k}}\det\left(\epsilon\boldsymbol{I}+\boldsymbol{B}_{\boldsymbol{s}}\boldsymbol{B}_{\boldsymbol{s}}^{*}\right)\nonumber \\
 & \text{ }\text{ }=\sum_{l=0}^{m}{n-l \choose k-l}{m \choose l}\epsilon^{m-l}=\sum_{l=0}^{m}{n-m+l \choose k-m+l}{m \choose l}\epsilon^{l}.\label{eq:bound2}
\end{align}
Observe that
\begin{align*}
\frac{{n-m+l \choose k-m+l}}{{n-m \choose k-m}} & =\frac{\prod_{i=1}^{l}\left(n-m+i\right)}{\prod_{i=1}^{l}\left(k-m+i\right)}\leq\frac{\prod_{i=1}^{l}\left(n-m+i\right)}{l!}\\
 & ={n-m+l \choose l}\leq{n \choose l}.
\end{align*}
This taken collectively with (\ref{eq:bound2}) suggests that 
\begin{align}
 & {n \choose k}\min_{\boldsymbol{s}\in{[n] \choose k}}\det\left(\epsilon\boldsymbol{I}+\boldsymbol{B}_{\boldsymbol{s}}\boldsymbol{B}_{\boldsymbol{s}}^{*}\right)\leq{n-m \choose k-m}\sum_{l=0}^{m}{n \choose l}{m \choose l}\epsilon^{l}\nonumber \\
 & \text{ }\text{ }\leq{n-m \choose k-m}\sum_{l=0}^{m}{n+m \choose 2l}\left(\sqrt{\epsilon}\right)^{2l}\nonumber \\
 & \quad\leq{n-m \choose k-m}\left(1+\sqrt{\epsilon}\right)^{n+m}.\label{eq:5}
\end{align}

On the other hand, we claim that
\begin{equation}
{n-m+l \choose k-m+l}{m \choose l}\leq{n-m \choose k-m}{K \choose l},\quad0\leq l\leq m\label{eq:claim_UB}
\end{equation}
for some integer $K\geq m$. Putting this claim and (\ref{eq:bound2})
together leads to
\begin{align*}
{n \choose k}\min_{\boldsymbol{s}\in{[n] \choose k}}\det\left(\epsilon\boldsymbol{I}+\boldsymbol{B}_{\boldsymbol{s}}\boldsymbol{B}_{\boldsymbol{s}}^{*}\right) & \leq\sum_{l=0}^{m}{n-m \choose k-m}{K \choose l}\epsilon^{l}\\
 & \leq{n-m \choose k-m}\left(1+\epsilon\right)^{K}.
\end{align*}
This together with (\ref{eq:5}) implies that
\begin{align}
 & \min_{\boldsymbol{s}\in{[n] \choose k}}\frac{1}{n}\log\det\left(\epsilon\boldsymbol{I}+\boldsymbol{B}_{\boldsymbol{s}}\boldsymbol{B}_{\boldsymbol{s}}^{*}\right)\nonumber \\
 & \quad\leq\frac{1}{n}\log{n-m \choose k-m}-\frac{1}{n}\log{n \choose k}+\nonumber \\
 & \quad\quad\quad\underset{:=\varrho_{2}\left(\epsilon\right)}{\underbrace{\min\left\{ \frac{K}{n}\log\left(1+\epsilon\right),\frac{n+m}{n}\log\left(1+\sqrt{\epsilon}\right)\right\} }}\\
 & \quad\leq\left(1-\alpha\right)\mathcal{H}\left(\frac{\beta-\alpha}{1-\alpha}\right)-\mathcal{H}\left(\beta\right)+\frac{\log\left(n+1\right)}{n}+\varrho_{2}\left(\epsilon\right)\nonumber \\
 & \quad=-\mathcal{H}\left(\alpha\right)+\beta\mathcal{H}\left(\frac{\alpha}{\beta}\right)+\frac{\log\left(n+1\right)}{n}+\varrho_{2}\left(\epsilon\right),\label{eq:log-UB-sub-Landau}
\end{align}
where the last inequality results from the fact $\log\left(1+\epsilon\right)\leq\epsilon$
as well as the following identity:
\begin{align*}
 & \frac{1}{n}\log\frac{{n-m \choose k-m}}{{n \choose k}}=\left(1-\alpha\right)\mathcal{H}\left(\frac{\beta-\alpha}{1-\alpha}\right)-\mathcal{H}\left(\beta\right)\\
 & \quad=-\left(\beta-\alpha\right)\log\left(\frac{\beta-\alpha}{1-\alpha}\right)-\left(1-\beta\right)\log\left(\frac{1-\beta}{1-\alpha}\right)-\mathcal{H}\left(\beta\right)\\
 & \quad=-\left(\beta-\alpha\right)\log\left(\beta-\alpha\right)+\left(1-\alpha\right)\log\left(1-\alpha\right)+\beta\log\beta\\
 & \quad=-\mathcal{H}\left(\alpha\right)+\beta\mathcal{H}\left(\frac{\alpha}{\beta}\right).
\end{align*}

Finally, it remains to establish the claim (\ref{eq:claim_UB}). In
fact, when $K\geq m$, one has 
\begin{align*}
\frac{{n-m \choose k-m}{K \choose l}}{{n-m+l \choose k-m+l}{m \choose l}} & =\frac{\prod_{i=0}^{l-1}\left(K-i\right)}{\prod_{i=0}^{l-1}\left(m-i\right)}\cdot\frac{\prod_{i=1}^{l}\left(k-m+i\right)}{\prod_{i=1}^{l}\left(n-m+i\right)}\\
 & \geq\left(\frac{K}{m}\right)^{l}\cdot\left(\frac{k-m}{n-m}\right)^{l}\geq1,
\end{align*}
provided that $K\geq\frac{n-m}{k-m}m=\frac{1-\alpha}{\beta-\alpha}\cdot m$,
which justifies the claim (\ref{eq:claim_UB}). This taken collectively
with (\ref{eq:log-UB-sub-Landau}) leads to
\begin{align*}
 & \min_{\boldsymbol{s}\in{[n] \choose k}}\frac{1}{n}\log\det\left(\epsilon\boldsymbol{I}+\boldsymbol{B}_{\boldsymbol{s}}\boldsymbol{B}_{\boldsymbol{s}}^{*}\right)\\
 & \quad\leq-\mathcal{H}\left(\alpha\right)+\beta\mathcal{H}\left(\frac{\alpha}{\beta}\right)+\frac{\log\left(n+1\right)}{n}\\
 & \quad\quad+\min\left\{ \left\lceil \frac{1-\alpha}{\beta-\alpha}\right\rceil \log\left(1+\epsilon\right),\left(1+\alpha\right)\log\left(1+\sqrt{\epsilon}\right)\right\} ,
\end{align*}
concluding the proof.

\section{Proof of Theorem \ref{theorem-Minimax-RMT-SuperLandau} \label{sec:Proof-of-theorem-Minimax-RMT-SuperLandau}}

\subsection{Proof of Theorem \ref{theorem-Minimax-RMT-SuperLandau}(a)}

Our goal is to evaluate $\frac{1}{n}\log\det\left(\epsilon\boldsymbol{I}_{k}+\boldsymbol{M}_{\boldsymbol{s}}^{\top}\left(\boldsymbol{M}\boldsymbol{M}^{\top}\right)^{-1}\boldsymbol{M}_{\boldsymbol{s}}\right)$
for some small $\epsilon>0$. We first define two Wishart matrices
\begin{align}
\Xi_{\backslash\boldsymbol{s}} & :=\frac{1}{n}\boldsymbol{M}\boldsymbol{M}^{\top}-\frac{1}{n}\boldsymbol{M}_{\boldsymbol{s}}\boldsymbol{M}_{\boldsymbol{s}}^{\top};\\
\Xi_{\boldsymbol{s}} & :=\frac{1}{n}\boldsymbol{M}_{\boldsymbol{s}}\boldsymbol{M}_{\boldsymbol{s}}^{\top}.
\end{align}
Apparently, $\Xi_{\boldsymbol{s}}\sim\mathcal{W}_{m}\left(k,\frac{1}{n}\boldsymbol{I}_{m}\right)$
and $\Xi_{\backslash\boldsymbol{s}}\sim\mathcal{W}_{m}\left(n-k,\frac{1}{n}\boldsymbol{I}_{m}\right)$.
When $1-\alpha>\beta$, i.e. $n-k>m$, $ $the Wishart matrix $\Xi_{\backslash\boldsymbol{s}}$
is invertible with probability 1. 

One difficulty in evaluating $\det\left(\epsilon\boldsymbol{I}_{k}+\boldsymbol{M}_{\boldsymbol{s}}^{\top}\left(\boldsymbol{M}\boldsymbol{M}^{\top}\right)^{-1}\boldsymbol{M}_{\boldsymbol{s}}\right)$
is that $\boldsymbol{M}_{\boldsymbol{s}}$ and $\boldsymbol{M}\boldsymbol{M}^{\top}$
are not independent. This motivates us to decouple them first as follows
\begin{align*}
 & \small\det\left(\epsilon\boldsymbol{I}_{k}+\boldsymbol{M}_{\boldsymbol{s}}^{\top}\left(\boldsymbol{M}\boldsymbol{M}^{\top}\right)^{-1}\boldsymbol{M}_{\boldsymbol{s}}\right)\\
 & \text{ }\small=\epsilon^{k-m}\det\left(\epsilon\boldsymbol{I}_{m}+\left(\frac{1}{n}\boldsymbol{M}\boldsymbol{M}^{\top}\right)^{-1}\frac{1}{n}\boldsymbol{M}_{\boldsymbol{s}}\boldsymbol{M}_{\boldsymbol{s}}^{\top}\right)\\
 & \text{ }\small=\epsilon^{k-m}\det\left(\epsilon\frac{1}{n}\boldsymbol{M}\boldsymbol{M}^{\top}+\frac{1}{n}\boldsymbol{M}_{\boldsymbol{s}}\boldsymbol{M}_{\boldsymbol{s}}^{\top}\right)\det\left(\frac{1}{n}\boldsymbol{M}\boldsymbol{M}^{\top}\right)^{-1}\\
 & \text{ }\small=\epsilon^{k-m}\det\left(\epsilon\Xi_{\backslash\boldsymbol{s}}+\left(1+\epsilon\right)\Xi_{\boldsymbol{s}}\right)\det\left(\frac{1}{n}\boldsymbol{M}\boldsymbol{M}^{\top}\right)^{-1}\\
 & \text{ }\small=\epsilon^{k-m}\det\left(\epsilon\boldsymbol{I}_{m}+\left(1+\epsilon\right)\Xi_{\boldsymbol{s}}\Xi_{\backslash\boldsymbol{s}}^{-1}\right)\det\left(\Xi_{\backslash\boldsymbol{s}}\right)\det\left(\frac{1}{n}\boldsymbol{M}\boldsymbol{M}^{\top}\right)^{-1}\\
 & \text{ }\small=\det\left(\epsilon\boldsymbol{I}_{k}+\frac{1+\epsilon}{n}\boldsymbol{M}_{\boldsymbol{s}}^{\top}\Xi_{\backslash\boldsymbol{s}}^{-1}\boldsymbol{M}_{\boldsymbol{s}}\right)\det\left(\Xi_{\backslash\boldsymbol{s}}\right)\det\left(\frac{1}{n}\boldsymbol{M}\boldsymbol{M}^{\top}\right)^{-1}
\end{align*}
or, equivalently,
\begin{align}
 & \frac{1}{n}\log\det\left(\epsilon\boldsymbol{I}_{k}+\boldsymbol{M}_{\boldsymbol{s}}^{\top}\left(\boldsymbol{M}\boldsymbol{M}^{\top}\right)^{-1}\boldsymbol{M}_{\boldsymbol{s}}\right)\nonumber \\
 & \quad=\frac{1}{n}\log\det\left(\epsilon\boldsymbol{I}_{k}+\left(1+\epsilon\right)\frac{1}{n}\boldsymbol{M}_{\boldsymbol{s}}^{\top}\Xi_{\backslash\boldsymbol{s}}^{-1}\boldsymbol{M}_{\boldsymbol{s}}\right)\nonumber \\
 & \quad\text{ }\quad+\frac{1}{n}\log\det\left(\Xi_{\backslash\boldsymbol{s}}\right)-\frac{1}{n}\log\det\left(\frac{1}{n}\boldsymbol{M}\boldsymbol{M}^{\top}\right).\label{eq:logdet-identity-separate-New}
\end{align}
The point of developing this identity (\ref{eq:logdet-identity-separate-New})
is to decouple the left-hand side of (\ref{eq:logdet-identity-separate-New})
through 3 matrices $\boldsymbol{M}_{\boldsymbol{s}}^{\top}\Xi_{\backslash\boldsymbol{s}}^{-1}\boldsymbol{M}_{\boldsymbol{s}}$,
$\Xi_{\backslash\boldsymbol{s}}$ and $\boldsymbol{M}\boldsymbol{M}^{\top}$.
In particular, since $\boldsymbol{M}_{\boldsymbol{s}}$ and $\Xi_{\backslash\boldsymbol{s}}$
are jointly independent, we can examine the concentration of measure
for $\boldsymbol{M}_{\boldsymbol{s}}$ and $\Xi_{\backslash\boldsymbol{s}}$
separately when evaluating $\boldsymbol{M}_{\boldsymbol{s}}^{\top}\Xi_{\backslash\boldsymbol{s}}^{-1}\boldsymbol{M}_{\boldsymbol{s}}$. 

The second and third terms of (\ref{eq:logdet-identity-separate-New})
can be evaluated through Lemma \ref{lem:GaussianLogDet_LB}. Specifically,
Lemma \ref{lem:GaussianLogDet_LB} indicates that
\begin{equation}
\frac{1}{n}\log\det\left(\frac{1}{n}\boldsymbol{M}\boldsymbol{M}^{\top}\right)\leq-\left(1-\alpha\right)\log\left(1-\alpha\right)-\alpha+\mathcal{O}\left(\frac{1}{\sqrt{n}}\right)\label{eq:LogDetMMTSuperLandau}
\end{equation}
with probability at least $1-C_{6}\exp\left(-2n\right)$ for some
constant $C_{6}>0$, and that for all $\boldsymbol{s}\in{[n] \choose k}$,
\begin{align}
 & \frac{1}{n}\log\det\left(\Xi_{\backslash\boldsymbol{s}}\right)=\frac{\log\det\left(\frac{n}{n-k}\Xi_{\backslash\boldsymbol{s}}\right)}{n}+\frac{\log\det\left(\frac{n-k}{n}\boldsymbol{I}\right)}{n}\nonumber \\
 & \quad\geq\left(1-\beta\right)\left\{ -\left(1-\frac{\alpha}{1-\beta}\right)\log\left(1-\frac{\alpha}{1-\beta}\right)-\frac{\alpha}{1-\beta}\right\} \nonumber \\
 & \quad\quad\quad\quad+\alpha\log\left(1-\beta\right)+\mathcal{O}\left(\frac{\log n}{n^{1/3}}\right)\\
 & \quad\geq-\left(1-\alpha-\beta\right)\log\left(1-\frac{\alpha}{1-\beta}\right)-\alpha+\alpha\log\left(1-\beta\right)\nonumber \\
 & \quad\quad\quad\quad+\mathcal{O}\left(\frac{\log n}{n^{1/3}}\right).\label{eq:LogDetMMTNotSSuperLandau}
\end{align}
hold simultaneously with probability exceeding $1-C_{9}\exp\left(-2n\right)$. 

Our main task then amounts to quantifying $\log\det\left(\epsilon\boldsymbol{I}_{k}+\boldsymbol{M}_{\boldsymbol{s}}^{\top}\Xi_{\backslash\boldsymbol{s}}^{-1}\boldsymbol{M}_{\boldsymbol{s}}\right)$,
which can be lower bounded via Lemma \ref{lemma-Expected-Log-Determinant-LambdaStatistics}.
This together with (\ref{eq:LogDetMMTSuperLandau}), (\ref{eq:LogDetMMTNotSSuperLandau})
and (\ref{eq:logdet-identity-separate-New}) yields that 
\begin{align}
 & \frac{1}{n}\log\det\left(\epsilon\boldsymbol{I}_{k}+\boldsymbol{M}_{\boldsymbol{s}}^{\top}\left(\boldsymbol{M}\boldsymbol{M}^{\top}\right)^{-1}\boldsymbol{M}_{\boldsymbol{s}}\right)\nonumber \\
 & \quad\geq-\left(\alpha-\beta\right)\log\left(\alpha-\beta\right)+\alpha\log\alpha\nonumber \\
 & \quad\quad+\left(1-\alpha-\beta\right)\log\left(1-\frac{\beta}{1-\alpha}\right)-\beta\log\left(1-\alpha\right)\nonumber \\
 & \quad\quad-\left(1-\alpha-\beta\right)\log\left(1-\frac{\alpha}{1-\beta}\right)-\alpha+\alpha\log\left(1-\beta\right)\nonumber \\
 & \quad\quad+\left(1-\alpha\right)\log\left(1-\alpha\right)+\alpha-\mathcal{O}\left(\frac{\log n}{n^{1/3}}\right)\\
 & \quad=\alpha\mathcal{H}\left(\frac{\beta}{\alpha}\right)-\mathcal{H}\left(\beta\right)-\mathcal{O}\left(\frac{\log n}{n^{1/3}}\right)
\end{align}
with probability exceeding $1-C_{9}\exp\left(-2n\right)$ for some
constants $C_{9}>0$. 

Since there are at most ${n \choose k}\leq2^{n}$ different states
$\boldsymbol{s}$, applying the union bound over all states completes
the proof.

\subsection{Proof of Theorem \ref{theorem-Minimax-RMT-SuperLandau}(b) }

We first recognize that
\begin{align}
 & \frac{1}{n}\log\det\left(\epsilon\boldsymbol{I}_{m}+(\boldsymbol{M}\boldsymbol{M}^{\top})^{-\frac{1}{2}}\boldsymbol{M}_{\boldsymbol{s}}\boldsymbol{M}_{\boldsymbol{s}}^{\top}(\boldsymbol{M}\boldsymbol{M}^{\top})^{-\frac{1}{2}}\right)\nonumber \\
 & \text{ }\small=-\frac{1}{n}\log\det\left(\frac{1}{n}\boldsymbol{M}\boldsymbol{M}^{\top}\right)+\frac{1}{n}\log\det\left(\frac{\epsilon\boldsymbol{M}\boldsymbol{M}^{\top}+\boldsymbol{M}_{\boldsymbol{s}}\boldsymbol{M}_{\boldsymbol{s}}^{\top}}{n}\right)\nonumber \\
 & \text{ }\small\geq-\frac{1}{n}\log\det\left(\frac{1}{n}\boldsymbol{M}\boldsymbol{M}^{\top}\right)+\frac{1}{n}\log\det\left(\frac{1}{n}\boldsymbol{M}_{\boldsymbol{s}}\boldsymbol{M}_{\boldsymbol{s}}^{\top}\right)\nonumber \\
 & \text{ }\small=-\frac{1}{n}\log\det\left(\frac{1}{n}\boldsymbol{M}\boldsymbol{M}^{\top}\right)+\frac{\beta}{k}\log\det\left(\frac{1}{k}\boldsymbol{M}_{\boldsymbol{s}}\boldsymbol{M}_{\boldsymbol{s}}^{\top}\right)+\alpha\log\beta.
\end{align}
When $\alpha<\beta\leq1$, Lemma \ref{lem:GaussianLogDet_LB} implies
that
\begin{align}
\frac{1}{n}\log\det\left(\frac{1}{n}\boldsymbol{M}\boldsymbol{M}^{\top}\right) & \leq\left(1-\alpha\right)\log\frac{1}{1-\alpha}-\alpha+\frac{c_{8}}{\sqrt{n}}
\end{align}
and
\begin{align}
\frac{1}{k}\log\det\left(\frac{1}{k}\boldsymbol{M}_{\boldsymbol{s}}\boldsymbol{M}_{\boldsymbol{s}}^{\top}\right) & \geq\left(1-\frac{\alpha}{\beta}\right)\log\frac{1}{1-\frac{\alpha}{\beta}}-\frac{\alpha}{\beta}-\frac{c_{9}\log n}{n^{1/3}}
\end{align}
hold with probability exceeding $1-9\exp\left(-3n\right)$, where
$c_{8}$ and $c_{9}$ are some positive universal constants. Putting
the above three bounds together gives
\begin{align}
 & \frac{1}{n}\log\det\left(\epsilon\boldsymbol{I}_{m}+(\boldsymbol{M}\boldsymbol{M}^{\top})^{-\frac{1}{2}}\boldsymbol{M}_{\boldsymbol{s}}\boldsymbol{M}_{\boldsymbol{s}}^{\top}(\boldsymbol{M}\boldsymbol{M}^{\top})^{-\frac{1}{2}}\right)\nonumber \\
 & \quad\geq-\left(1-\alpha\right)\log\frac{1}{1-\alpha}+\alpha+\left(\beta-\alpha\right)\log\frac{1}{1-\frac{\alpha}{\beta}}\nonumber \\
 & \quad\quad\quad-\alpha-\frac{c_{10}\log n}{n^{1/3}}+\alpha\log\beta\\
 & \quad=-\mathcal{H}\left(\alpha\right)+\beta\mathcal{H}\left(\frac{\alpha}{\beta}\right)-\frac{c_{10}\log n}{n^{1/3}}
\end{align}
for some universal constant $c_{10}>0$, where the last identity follows
since
\begin{align}
 & -\left(1-\alpha\right)\log\frac{1}{1-\alpha}+\left(\beta-\alpha\right)\log\frac{1}{1-\frac{\alpha}{\beta}}+\alpha\log\beta\nonumber \\
 & \quad=-\mathcal{H}\left(\alpha\right)-\alpha\log\alpha+\beta\mathcal{H}\left(\frac{\alpha}{\beta}\right)-\alpha\log\frac{\beta}{\alpha}+\alpha\log\beta\nonumber \\
 & \quad=-\mathcal{H}\left(\alpha\right)+\beta\mathcal{H}\left(\frac{\alpha}{\beta}\right).
\end{align}
Applying the union bound over all ${n \choose k}$ states concludes
the proof.

\subsection{Proof of Theorem \ref{theorem-Minimax-RMT-SuperLandau}(c) }

Without loss of generality, consider first the case where $\boldsymbol{s}=\left\{ n-k+1,\cdots,n\right\} $.
The quantity under study can be rearranged as
\begin{align}
 & \log\det\left(\epsilon\boldsymbol{I}_{k}+\boldsymbol{M}_{\boldsymbol{s}}^{\top}\left(\boldsymbol{M}\boldsymbol{M}^{\top}\right)^{-1}\boldsymbol{M}_{\boldsymbol{s}}\right)\nonumber \\
 & \quad=\log\left\{ \epsilon^{k-m}\det\left(\epsilon\boldsymbol{I}_{m}+\boldsymbol{M}_{\boldsymbol{s}}\boldsymbol{M}_{\boldsymbol{s}}^{\top}\left(\boldsymbol{M}\boldsymbol{M}^{\top}\right)^{-1}\right)\right\} \nonumber \\
 & \quad=\log\det\left(\epsilon\boldsymbol{M}\boldsymbol{M}^{\top}+\boldsymbol{M}_{\boldsymbol{s}}\boldsymbol{M}_{\boldsymbol{s}}^{\top}\right)-\log\det\left(\boldsymbol{M}\boldsymbol{M}^{\top}\right)\nonumber \\
 & \quad\quad\quad\quad+\left(k-m\right)\log\epsilon\\
 & \quad=\log\det\left(\frac{1}{n}\boldsymbol{M}\underset{:=\boldsymbol{D}_{\epsilon}}{\underbrace{\left[\begin{array}{cc}
\epsilon\boldsymbol{I}_{n-k}\\
 & \left(1+\epsilon\right)\boldsymbol{I}_{k}
\end{array}\right]}}\boldsymbol{M}^{\top}\right)\nonumber \\
 & \quad\quad\quad\quad-\log\det\left(\frac{1}{n}\boldsymbol{M}\boldsymbol{M}^{\top}\right)+\left(k-m\right)\log\epsilon.\label{eq:Rearrange}
\end{align}
The term $\log\det\left(\frac{1}{n}\boldsymbol{M}\boldsymbol{M}^{\top}\right)$
can be controlled by Lemma \ref{lem:GaussianLogDet_LB}. Thus, it
amounts to derive a reasonably tight lower bound on the term $\log\det\left(\frac{1}{n}\boldsymbol{M}\boldsymbol{D}_{\epsilon}\boldsymbol{M}^{\top}\right)$.

Fortunately, the concentration of spectral measure inequality \cite[Corollary 1.8]{GuionnetZeitouni2000}
can also be applied to control the quantity $\sum_{i=1}^{n}f\left(\lambda_{i}\left(\frac{1}{n}\boldsymbol{M}\boldsymbol{D}_{\epsilon}\boldsymbol{M}^{\top}\right)\right)$
for a variety of functions $f(\cdot)$. Consider the auxiliary functions
\begin{equation}
f_{1,\delta}\left(x\right):=\begin{cases}
\frac{2}{\sqrt{\delta}}\left(\sqrt{x}-\sqrt{\delta}\right)+\log\epsilon,\quad & 0<x<\delta,\\
\log x, & x\geq\delta,
\end{cases}\label{eq:Defn_f_1_epsilon-1}
\end{equation}
and
\begin{equation}
\text{det}^{\delta}\left(\boldsymbol{X}\right):=\prod_{i=1}^{m}e^{f_{1,\delta}\left(\lambda_{i}\left(\boldsymbol{X}\right)\right)}.\label{eq:Defn_Det_epsilon-1}
\end{equation}
If we set 
\[
Z_{\epsilon}:=\log\text{det}^{\delta}\left(\frac{1}{n}\boldsymbol{M}\boldsymbol{D}_{\epsilon}\boldsymbol{M}^{\top}\right)-\mathbb{E}\left[\log\text{det}^{\delta}\left(\frac{1}{n}\boldsymbol{M}\boldsymbol{D}_{\epsilon}\boldsymbol{M}^{\top}\right)\right],
\]
then using \cite[Corollary 1.8]{GuionnetZeitouni2000} we get\footnote{This arises since the concentration of spectral measure results given
in \cite[Corollary 1.8]{GuionnetZeitouni2000} depend only on the
spectral norm $\left\Vert \boldsymbol{D}_{\epsilon}\right\Vert $. }
\begin{equation}
\mathbb{P}\left\{ \left|\frac{1}{n}Z_{\epsilon}\right|>4\sqrt{\frac{\alpha\left(1+\epsilon\right)}{\delta}}\sqrt{\frac{\tau}{n}}\right\} <2\exp\left(-2\tau n\right)\label{eq:GaussianDetAA-1-epsilon}
\end{equation}
for any $\tau>0$. Furthermore, repeating the same argument as in
(\ref{eq:LowerBoundELogDetEpsilon}) and (\ref{eq:Lambda_min_AA})
we obtain 

\begin{align}
 & \frac{1}{n}\mathbb{E}\small\left[\log\text{det}^{\delta}\left(\frac{1}{n}\boldsymbol{M}\boldsymbol{D}_{\epsilon}\boldsymbol{M}^{\top}\right)\right]\geq\frac{1}{n}\log\mathbb{E}\small\left[\text{det}\left(\frac{1}{n}\boldsymbol{M}\boldsymbol{D}_{\epsilon}\boldsymbol{M}^{\top}\right)\right]\nonumber \\
 & \quad\quad\quad\quad\quad\quad\quad\quad\quad-\frac{5\alpha\left(1+\epsilon\right)}{n\delta},\quad\forall\delta<\left(1+\epsilon\right)\alpha,\label{eq:LowerBoundELogDetEpsilon-delta}
\end{align}
and for sufficiently large $n$, 
\begin{align}
\mathbb{P}\left\{ \lambda_{\min}\left(\frac{1}{n}\boldsymbol{M}\boldsymbol{D}_{\epsilon}\boldsymbol{M}^{\top}\right)<\frac{\epsilon}{n^{7/3}}\right\}  & <3e^{-3n}.
\end{align}
By setting $\delta=\epsilon^{2/3}n^{-1/3}$, one has, with probability
exceeding $1-7\exp\left(-\tau n\right)$, that
\begin{align}
 & \frac{1}{n}\log\det\left(\frac{1}{n}\boldsymbol{M}\boldsymbol{D}_{\epsilon}\boldsymbol{M}^{\top}\right)-\frac{1}{n}\log\mathrm{det}^{\delta}\left(\frac{1}{n}\boldsymbol{M}\boldsymbol{D}_{\epsilon}\boldsymbol{M}^{\top}\right)\nonumber \\
 & \text{ }\geq\text{ }\sum_{i:\lambda_{i}\left(\frac{1}{n}\boldsymbol{M}\boldsymbol{D}_{\epsilon}\boldsymbol{M}^{\top}\right)<\delta}\Big\{\log\lambda_{\min}\left(\frac{1}{n}\boldsymbol{M}\boldsymbol{D}_{\epsilon}\boldsymbol{M}^{\top}\right)-\log\delta\Big\}\nonumber \\
 & \text{ }\geq\text{ }-\frac{\text{card}\left\{ i\mid\lambda_{i}\left(\frac{\epsilon}{n}\boldsymbol{M}\boldsymbol{M}^{\top}\right)<\delta\right\} }{n}\cdot\log\left(\frac{n^{2}}{\epsilon^{1/3}}\right)\label{eq:second_inequality}\\
 & \text{ }\geq\text{ }-\left(\frac{\alpha}{\left(1-\alpha-\frac{1}{n}\right)}+4\sqrt{\epsilon\alpha\tau}\right)\cdot\frac{\log\left(\frac{n^{2}}{\epsilon^{1/3}}\right)}{\left(\epsilon n\right)^{1/3}},\label{eq:third_inequality}
\end{align}
where (\ref{eq:second_inequality}) follows since $\boldsymbol{M}\boldsymbol{D}_{\epsilon}\boldsymbol{M}^{\top}\succeq\epsilon\boldsymbol{M}\boldsymbol{M}^{\top}$,
and (\ref{eq:third_inequality}) arises from Lemma \ref{lem:GaussianLocalSpectrum}.
This combined with (\ref{eq:GaussianDetAA-1-epsilon}) and (\ref{eq:LowerBoundELogDetEpsilon})
suggests that for any $\epsilon\in\left(\frac{1}{n},1\right)$, there
exist universal constants $c_{1}\sim c_{4}>0$ such that
\begin{align}
 & \frac{1}{n}\log\det\left(\frac{1}{n}\boldsymbol{M}\boldsymbol{D}_{\epsilon}\boldsymbol{M}^{\top}\right)\nonumber \\
 & \text{ }\geq\text{ }\frac{1}{n}\mathbb{E}\left[\log\mathrm{det}^{\delta}\left(\frac{1}{n}\boldsymbol{M}\boldsymbol{D}_{\epsilon}\boldsymbol{M}^{\top}\right)\right]-\frac{\left(c_{1}+c_{2}\sqrt{\tau}\right)\log\frac{n^{2}}{\epsilon^{\frac{1}{3}}}}{\left(\epsilon n\right)^{1/3}}\label{eq:4th_inequality}\\
 & \text{ }\geq\text{ }\frac{1}{n}\log\mathbb{E}\left[\text{det}\left(\frac{1}{n}\boldsymbol{M}\boldsymbol{D}_{\epsilon}\boldsymbol{M}^{\top}\right)\right]-\frac{\left(c_{3}+c_{4}\sqrt{\tau}\right)\log n}{\left(\epsilon n\right)^{1/3}}\label{eq:5th_inequality}
\end{align}
with probability exceeding $1-7\exp\left(-\tau n\right)$.

It remains to develop a tight lower bound on $\frac{1}{n}\log\mathbb{E}\left[\text{det}\left(\frac{1}{n}\boldsymbol{M}\boldsymbol{D}_{\epsilon}\boldsymbol{M}^{\top}\right)\right]$.
Applying Cauchy-Binet identity gives
\begin{align}
 & \mathbb{E}\left[\text{det}\left(\frac{1}{n}\boldsymbol{M}\boldsymbol{D}_{\epsilon}\boldsymbol{M}^{\top}\right)\right]\nonumber \\
 & \quad=\sum_{\boldsymbol{r},\tilde{\boldsymbol{r}}\in{[n] \choose m}}\small\mathbb{E}\left[\det\left(\boldsymbol{M}_{\left[m\right],\boldsymbol{r}}\right)\det\left(\frac{1}{n}\left(\boldsymbol{D}_{\epsilon}\right)_{\boldsymbol{r},\tilde{\boldsymbol{r}}}\right)\det\left(\boldsymbol{M}_{\left[m\right],\tilde{\boldsymbol{r}}}\right)\right]\nonumber \\
 & \quad=\sum_{\boldsymbol{r}\in{[n] \choose m}}\mathbb{E}\left[\det\left(\boldsymbol{M}_{\left[m\right],\boldsymbol{r}}\boldsymbol{M}_{\left[m\right],\boldsymbol{r}}^{\top}\right)\right]\det\left(\frac{1}{n}\left(\boldsymbol{D}_{\epsilon}\right)_{\boldsymbol{r},\tilde{\boldsymbol{r}}}\right)\label{eq:D_diagonal_det}\\
 & \quad=\mathbb{E}\left[\det\left(\boldsymbol{M}_{\left[m\right],\boldsymbol{r}}\boldsymbol{M}_{\left[m\right],\boldsymbol{r}}^{\top}\right)\right]\sum_{\boldsymbol{r}\in{[n] \choose m}}\frac{1}{n^{m}}\det\left(\left(\boldsymbol{D}_{\epsilon}\right)_{\boldsymbol{r},\tilde{\boldsymbol{r}}}\right)\nonumber \\
 & \quad=\text{ }\frac{m!}{n^{m}}\sum_{l=\max\left\{ m-k,\text{ }0\right\} }^{\min\left\{ n-k,\text{ }m\right\} }\varphi_{n,k,m}\left(l\right),\label{eq:DetD_sum}
\end{align}
where we define $\varphi_{n,k,m}\left(l\right):={n-k \choose l}{k \choose m-l}\epsilon^{l}\left(1+\epsilon\right)^{m-l}$.
In the above arguments, the identity (\ref{eq:D_diagonal_det}) arises
since $\det\left(\left(\boldsymbol{D}_{\epsilon}\right)_{\boldsymbol{r},\tilde{\boldsymbol{r}}}\right)\neq0$
only if $\boldsymbol{r}=\tilde{\boldsymbol{r}}$. The identity (\ref{eq:DetD_sum})
follows from the definition of $\boldsymbol{D}_{\epsilon}$ as well
as the fact that $\mathbb{E}\left[\det\left(\boldsymbol{M}_{\left[m\right],\boldsymbol{r}}\boldsymbol{M}_{\left[m\right],\boldsymbol{r}}^{\top}\right)\right]=m!$
(e.g. \cite[Theorem 3.1]{edelman1997probability}). 

Since $m\geq k$, (\ref{eq:DetD_sum}) further yields
\begin{align*}
\mathbb{E}\left[\text{det}\left(\frac{1}{n}\boldsymbol{M}\boldsymbol{D}_{\epsilon}\boldsymbol{M}^{\top}\right)\right] & \geq\text{ }\frac{m!}{n^{m}}\varphi_{n,k,m}\left(m-k\right)\text{ }\\
 & =\text{ }\frac{m!}{n^{m}}{n-k \choose m-k}\epsilon^{m-k}\left(1+\epsilon\right)^{k}.
\end{align*}
As a result,
\begin{align*}
 & \frac{1}{n}\log\mathbb{E}\left[\text{det}\left(\frac{1}{n}\boldsymbol{M}\boldsymbol{D}_{\epsilon}\boldsymbol{M}^{\top}\right)\right]\\
 & \text{ }\geq\frac{1}{n}\log\left\{ \frac{m!}{n^{m}}{n-k \choose m-k}\epsilon^{m-k}\right\} \\
 & \text{ }=\frac{\log m!}{n}-\frac{m\log m}{n}+\frac{m\log\left(\frac{m}{n}\right)+\log{n-k \choose m-k}+\log\epsilon^{m-k}}{n}\\
 & \text{ }\geq-\alpha+\alpha\log\alpha+\left(1-\beta\right)\mathcal{H}\left(\frac{\alpha-\beta}{1-\beta}\right)+\frac{\log\epsilon^{m-k}}{n}-\frac{\log n}{n}.
\end{align*}
Here, the last inequality makes use of the entropy formula (\ref{eq:EntropyApproximation})
as well as the following inequality
\begin{align}
\frac{\log m!}{n}-\frac{m\log m}{n} & \geq\frac{\left(m+\frac{1}{2}\right)}{n}\log m-\frac{m}{n}-\frac{m}{n}\log m\nonumber \\
 & \geq-\alpha,\label{eq:identity1}
\end{align}
a consequence of the Stirling approximation $ $(\ref{eq:Stirling_LB}).
Substituting it back into the concentration bound (\ref{eq:5th_inequality})
we get
\begin{align}
 & \frac{1}{n}\log\det\left(\frac{1}{n}\boldsymbol{M}\boldsymbol{D}_{\epsilon}\boldsymbol{M}^{\top}\right)\nonumber \\
 & \quad\geq-\alpha+\alpha\log\alpha+\left(1-\beta\right)\mathcal{H}\left(\frac{\alpha-\beta}{1-\beta}\right)-\frac{\log n}{n}\nonumber \\
 & \quad\quad-\frac{\left(c_{3}+c_{4}\sqrt{\tau}\right)\log n}{\left(\epsilon n\right)^{1/3}}+\frac{1}{n}\log\left(\epsilon^{m-k}\right)
\end{align}
with probability at least $1-7\exp\left(-\tau n\right)$.

So far we have developed lower bounds on the term $\frac{1}{n}\log\det\left(\frac{1}{n}\boldsymbol{M}\boldsymbol{D}_{\epsilon}\boldsymbol{M}^{\top}\right)$.
The above bound taken collectively with (\ref{eq:Rearrange}) and
Lemma \ref{lem:GaussianLogDet_LB} leads to
\begin{align}
 & \frac{1}{n}\log\det\left(\epsilon\boldsymbol{I}_{k}+\boldsymbol{M}_{\boldsymbol{s}}^{\top}\left(\boldsymbol{M}\boldsymbol{M}^{\top}\right)^{-1}\boldsymbol{M}_{\boldsymbol{s}}\right)\nonumber \\
 & \text{ }=\frac{\log\det\left(\frac{1}{n}\boldsymbol{M}\boldsymbol{D}_{\epsilon}\boldsymbol{M}^{\top}\right)-\log\det\left(\frac{1}{n}\boldsymbol{M}\boldsymbol{M}^{\top}\right)}{n}+\frac{k-m}{n}\log\epsilon\nonumber \\
 & \text{ }\geq-\alpha+\alpha\log\alpha+\left(1-\beta\right)\mathcal{H}\left(\frac{\alpha-\beta}{1-\beta}\right)-\frac{\left(c_{3}+c_{4}\sqrt{\tau}\right)\log n}{\left(\epsilon n\right)^{1/3}}\nonumber \\
 & \quad\quad-\frac{\log n}{n}-\left\{ \left(1-\alpha\right)\log\frac{1}{1-\alpha}-\alpha+\frac{2\log n}{n}+c_{5}\sqrt{\frac{\tau}{n}}\right\} \nonumber \\
 & \text{ }\geq\text{ }\alpha\mathcal{H}\left(\frac{\beta}{\alpha}\right)-\mathcal{H}\left(\beta\right)-\frac{\left(c_{5}+c_{6}\sqrt{\tau}\right)\log n}{\left(\epsilon n\right)^{1/3}}\label{eq:last_ineq}
\end{align}
with probability at least $1-9\exp\left(-\tau n\right)$, where $c_{5},c_{6}>0$
are some universal constants. Here, the inequality (\ref{eq:last_ineq})
follows from the following identity:
\begin{align}
 & -\alpha+\alpha\log\alpha+\left(1-\beta\right)\mathcal{H}\left(\frac{\alpha-\beta}{1-\beta}\right)-\left(1-\alpha\right)\log\frac{1}{1-\alpha}+\alpha\nonumber \\
 & \quad=-\alpha+\alpha\log\alpha-\left(\alpha-\beta\right)\log\left(\frac{\alpha-\beta}{1-\beta}\right)+\alpha\nonumber \\
 & \quad\quad\quad-\left(1-\alpha\right)\log\left(\frac{1-\alpha}{1-\beta}\right)+\left(1-\alpha\right)\log\left(1-\alpha\right)\nonumber \\
 & \quad=\alpha\log\alpha-\left(\alpha-\beta\right)\log\left(\alpha-\beta\right)+\left(1-\beta\right)\log\left(1-\beta\right)\nonumber \\
 & \quad=\alpha\mathcal{H}\left(\frac{\beta}{\alpha}\right)-\mathcal{H}\left(\beta\right).
\end{align}
The proof is then complete by applying the union bound and setting
$\tau=3$.

\section{Proof of Lemma \ref{lem:GaussianLocalSpectrum}\label{sec:Proof-of-Lemma-Gaussian-Local-Spectral}}

Since the indicator function ${\bf 1}_{[0,\delta]}\left(\cdot\right)$
entails discontinuous points, we consider instead an upper bound on
${\bf 1}_{[0,\delta]}(\cdot)$ as
\begin{equation}
f_{2,\delta}(x):=\begin{cases}
1, & \text{if }0\leq x\leq\delta;\\
-x/\epsilon+2,\quad & \text{if }\delta<x\leq2\delta;\\
0, & \text{else}.
\end{cases}
\end{equation}
Note that for any $\delta>0$ and $x\geq0$, one has
\[
\frac{\delta}{x}-\left(-\frac{x}{\delta}+2\right)=\frac{\delta}{x}+\frac{x}{\delta}-2\geq2\sqrt{\frac{\delta}{x}\cdot\frac{x}{\delta}}-2=0.
\]
This together with the facts $\frac{\epsilon}{x}\geq1$ ($0\leq x\leq\epsilon$)
and $\frac{\epsilon}{x}\geq0$ ($x\geq0$) indicates that
\begin{equation}
f_{2,\delta}(x)\leq\frac{\delta}{x},\quad\forall x\geq0,
\end{equation}
leading to the upper bound
\begin{align}
\sum_{i=1}^{n}f_{2,\delta}\left(\lambda_{i}\left(\frac{1}{n}\boldsymbol{A}\boldsymbol{A}^{\top}\right)\right) & \leq\sum_{i=1}^{m}\frac{\delta}{\lambda_{i}\left(\frac{1}{n}\boldsymbol{A}\boldsymbol{A}^{\top}\right)}\nonumber \\
 & =\delta\text{tr}\left(\left(\frac{1}{n}\boldsymbol{A}\boldsymbol{A}^{\top}\right)^{-1}\right).
\end{align}
It then follows from the property of inverse Wishart matrices (e.g.
\cite[Theorem 2.2.8]{Fujikoshi2010}) that
\begin{align}
 & \mathbb{E}\left[\frac{1}{n}\sum_{i=1}^{n}f_{2,\delta}\left(\lambda_{i}\left(\frac{1}{n}\boldsymbol{A}\boldsymbol{A}^{\top}\right)\right)\right]\leq\frac{\delta}{n}\mathbb{E}\left[\text{tr}\left(\left(\frac{1}{n}\boldsymbol{A}\boldsymbol{A}^{\top}\right)^{-1}\right)\right]\nonumber \\
 & \quad=\frac{m}{n-m-1}\delta=\frac{\alpha}{1-\alpha-\frac{1}{n}}\delta.
\end{align}

Clearly, the Lipschitz constant of the function 
\[
g_{2,\delta}(x):=f_{2,\delta}(x^{2})=\begin{cases}
1, & \text{if }0\leq x\leq\sqrt{\delta};\\
-x^{2}/\delta+2,\quad & \text{if }\sqrt{\delta}<x\leq\sqrt{2\delta}\\
0, & \text{else}
\end{cases};
\]
 is bounded above by $\sqrt{8/\delta}$. Applying \cite[Corollary 1.8(b)]{GuionnetZeitouni2000}
then yields the following: for any $\varepsilon>0$,
\begin{align}
 & \mathbb{P}\left(\frac{1}{n}\sum_{i=1}^{n}{\bf 1}_{[0,\delta]}\left(\lambda_{i}\left(\frac{1}{n}\boldsymbol{A}\boldsymbol{A}^{\top}\right)\right)>\frac{\alpha}{1-\alpha-\frac{1}{n}}\delta+\varepsilon\right)\nonumber \\
 & \quad\leq\text{ }\mathbb{P}\left(\frac{1}{n}\sum_{i=1}^{n}f_{2,\delta}\left(\lambda_{i}\left(\frac{1}{n}\boldsymbol{A}\boldsymbol{A}^{\top}\right)\right)>\frac{\alpha}{1-\alpha-\frac{1}{n}}\delta+\varepsilon\right)\nonumber \\
 & \quad\leq\text{ }\mathbb{P}\left(\frac{1}{n}\sum_{i=1}^{n}f_{2,\delta}\left(\lambda_{i}\left(\frac{1}{n}\boldsymbol{A}\boldsymbol{A}^{\top}\right)\right)>\right.\nonumber \\
 & \quad\quad\quad\quad\quad\left.\mathbb{E}\left[\frac{1}{n}\sum_{i=1}^{n}f_{2,\delta}\left(\lambda_{i}\left(\frac{1}{n}\boldsymbol{A}\boldsymbol{A}^{\top}\right)\right)\right]+\varepsilon\right)\nonumber \\
 & \quad\leq\text{ }2\exp\left(-\frac{\delta}{16\alpha}\varepsilon^{2}n^{2}\right).
\end{align}
Put in other words, for any $\tau>0$, one has
\begin{equation}
\frac{\text{card}\left\{ i\mid\lambda_{i}\left(\frac{1}{n}\boldsymbol{A}\boldsymbol{A}^{\top}\right)<\delta\right\} }{n}<\frac{\alpha}{1-\alpha-\frac{1}{n}}\delta+\frac{4\sqrt{\alpha\tau}}{\sqrt{n\delta}}\label{eq:SmallEvalueCardinalityBound}
\end{equation}
with probability exceeding $1-2\exp\left(-\tau n\right)$, as claimed. 

In particular, by setting $\delta=n^{-1/3}$, one has
\begin{equation}
\frac{\text{card}\left\{ i\mid\lambda_{i}\left(\frac{1}{n}\boldsymbol{A}\boldsymbol{A}^{\top}\right)<\frac{1}{n^{1/3}}\right\} }{n}<\frac{\frac{\alpha}{1-\alpha-\frac{1}{n}}+4\sqrt{\alpha\tau}}{n^{1/3}}\label{eq:CardinalityEpsilon}
\end{equation}
with probability at least $1-2\exp\left(-\tau n\right)$.

\section{Proof of Lemma \ref{lem:GaussianLogDet_LB}\label{sec:Proof-of-Lemma-GaussianLogDet_LB}}

Before proceeding to the proof of measure concentration, we first
derive tight bounds on the quantity $\frac{1}{n}\log\mathbb{E}\left[\det\left(\frac{1}{n}\boldsymbol{A}\boldsymbol{A}^{\top}\right)\right]$.
First, the Cauchy-Binet formula indicates that
\begin{equation}
\mathbb{E}\left[\det\left(\boldsymbol{A}\boldsymbol{A}^{\top}\right)\right]=\sum_{\boldsymbol{s}\in{[n] \choose m}}\mathbb{E}\left[\det\left(\boldsymbol{A}_{\boldsymbol{s}}\boldsymbol{A}_{\boldsymbol{s}}^{\top}\right)\right],
\end{equation}
where $\boldsymbol{s}$ ranges over all $m$-combinations of $\{1,\cdots,n\}$,
and $\boldsymbol{A}_{\boldsymbol{s}}$ is the $m\times m$ minor of
$\boldsymbol{A}$ whose columns come from the columns of $\boldsymbol{A}$
at indices from ${\bf s}$. It is well known that (e.g. \cite[Theorem 3.1]{edelman1997probability})
for i.i.d. random ensembles with zero mean and unit variance, the
determinant satisfies
\begin{equation}
\mathbb{E}\left[\det\left(\boldsymbol{A}_{\boldsymbol{s}}\boldsymbol{A}_{\boldsymbol{s}}^{\top}\right)\right]=m!,\label{eq:E_det_AA}
\end{equation}
which immediately leads to
\begin{align}
\mathbb{E}\left[\det\left(\frac{1}{n}\boldsymbol{A}\boldsymbol{A}^{\top}\right)\right] & =\frac{\sum_{\boldsymbol{s}\in{[n] \choose m}}\mathbb{E}\left[\det\left(\boldsymbol{A}_{\boldsymbol{s}}\boldsymbol{A}_{\boldsymbol{s}}^{\top}\right)\right]}{n^{m}}\nonumber \\
 & =\frac{m!}{n^{m}}{n \choose m}.\label{eq:E_det_expansion}
\end{align}
Next, from the well-known Stirling inequality
\begin{equation}
\sqrt{2\pi}m^{m+\frac{1}{2}}e^{-m}\leq m!\leq em^{m+\frac{1}{2}}e^{-m},
\end{equation}
one can obtain
\begin{equation}
\log\left(m!\right)\leq\left(m+\frac{1}{2}\right)\log m-m+1;
\end{equation}
\begin{equation}
\log\left(m!\right)\geq\left(m+\frac{1}{2}\right)\log m-m+\frac{1}{2}\log\left(2\pi\right).\label{eq:Stirling_LB}
\end{equation}
These together with the entropy formula (\ref{eq:EntropyApproximation})
give rise to
\begin{align}
 & \frac{1}{n}\log\mathbb{E}\left[\det\left(\frac{1}{n}\boldsymbol{A}\boldsymbol{A}^{\top}\right)\right]\nonumber \\
 & \quad\leq-\frac{m}{n}\log n+\frac{\left(m+\frac{1}{2}\right)\log m}{n}-\frac{m}{n}+\frac{1}{n}+\mathcal{H}\left(\frac{m}{n}\right)\nonumber \\
 & \quad=\left(1-\alpha\right)\log\frac{1}{1-\alpha}-\alpha+\frac{\log\left(e^{2}m\right)}{2n}\label{eq:UB_logEdet}
\end{align}
and, similarly, 
\begin{align}
 & \frac{1}{n}\log\mathbb{E}\left[\det\left(\frac{1}{n}\boldsymbol{A}\boldsymbol{A}^{\top}\right)\right]\nonumber \\
 & \quad\geq\left(1-\alpha\right)\log\frac{1}{1-\alpha}-\alpha-\frac{\log\left(n+1\right)}{n}.\label{eq:LB_logEdet}
\end{align}

(1) We are now in a position to establish the upper bound on $\log\det\left(\frac{1}{n}\boldsymbol{A}\boldsymbol{A}^{\top}\right)$.
Since $\log\det\left(\frac{1}{n}\boldsymbol{A}\boldsymbol{A}^{\top}\right)=\sum_{i=1}^{m}\log\left(\lambda_{i}\left(\frac{1}{n}\boldsymbol{A}\boldsymbol{A}^{\top}\right)\right)$
is a separable function on the spectrum of $\frac{1}{n}\boldsymbol{A}\boldsymbol{A}^{\top}$,
one strategy is to make use of the concentration of spectral measure
results \cite{GuionnetZeitouni2000}. Note, however, that the function
$\log x$ does not satisfy the Lipschitz condition required therein.
To resolve this issue, we define an auxiliary function
\begin{equation}
f_{1,\delta}\left(x\right):=\begin{cases}
\frac{2}{\sqrt{\delta}}\left(\sqrt{x}-\sqrt{\delta}\right)+\log\epsilon,\quad & 0<x<\delta,\\
\log x, & x\geq\delta,
\end{cases}\label{eq:Defn_f_1_epsilon}
\end{equation}
as well as
\begin{equation}
\text{det}^{\delta}\left(\boldsymbol{X}\right):=\prod_{i=1}^{m}e^{f_{1,\delta}\left(\lambda_{i}\left(\boldsymbol{X}\right)\right)}.\label{eq:Defn_Det_epsilon}
\end{equation}
Apparently, $f_{1,\delta}\left(x\right)\geq\log x$, and the Lipschitz
constant of the \emph{concave} function
\[
g_{1,\delta}\left(x\right):=f_{1,\delta}\left(x^{2}\right)=\begin{cases}
\frac{2}{\sqrt{\delta}}\left(x-\sqrt{\delta}\right)+\log\delta,\quad & 0<x<\sqrt{\delta},\\
2\log x & x\geq\sqrt{\delta},
\end{cases}
\]
is bounded above by $\frac{2}{\sqrt{\delta}}$. By definition, 
\begin{equation}
\text{det}^{\delta}\left(\frac{1}{n}\boldsymbol{A}\boldsymbol{A}^{\top}\right)=\text{det}\left(\frac{1}{n}\boldsymbol{A}\boldsymbol{A}^{\top}\right)
\end{equation}
holds in the event that $\left\{ \lambda_{\min}\left(\frac{1}{n}\boldsymbol{A}\boldsymbol{A}^{\top}\right)\geq\delta\right\} $.

The deviation of $\log\text{det}^{\delta}\left(\frac{1}{n}\boldsymbol{A}\boldsymbol{A}^{\top}\right)$
can be controlled via the concentration of spectral measure inequalities.
Specifically, if we set
\begin{equation}
Z:=\log\text{det}^{\delta}\left(\frac{1}{n}\boldsymbol{A}\boldsymbol{A}^{\top}\right)-\mathbb{E}\left[\log\text{det}^{\delta}\left(\frac{1}{n}\boldsymbol{A}\boldsymbol{A}^{\top}\right)\right],\label{eq:DefnZ}
\end{equation}
then \cite[Corollary 1.8]{GuionnetZeitouni2000} suggests that for
any $\tau>0$,
\begin{equation}
\mathbb{P}\left(\left|Z\right|>\tau\right)\leq2\exp\left(-\frac{\delta\tau^{2}}{8\alpha}\right)\label{eq:UB_Z}
\end{equation}
or, equivalently, 
\begin{equation}
\mathbb{P}\left\{ \left|\frac{1}{n}Z\right|>4\sqrt{\frac{\alpha}{\delta}}\sqrt{\frac{\tau}{n}}\right\} <2\exp\left(-2\tau n\right).\label{eq:GaussianDetAA-1}
\end{equation}

Since $\log\text{det}\left(\frac{1}{n}\boldsymbol{A}\boldsymbol{A}^{\top}\right)\leq\log\text{det}^{\delta}\left(\frac{1}{n}\boldsymbol{A}\boldsymbol{A}^{\top}\right)$,
it amounts to derive a tight upper bound on $\mathbb{E}\left[\log\text{det}^{\delta}\left(\frac{1}{n}\boldsymbol{A}\boldsymbol{A}^{\top}\right)\right]$.
Note that the behavior of the least singular value of a rectangular
Gaussian matrix has been largely studied in the random matrix literature
(e.g. \cite[Corollary 5.35]{Vershynin2012}), which we cite as follows.

\begin{lem}\label{lemma-Concentration-of-Norm-M}Consider a Gaussian
random matrix $\boldsymbol{M}\in\mathbb{R}^{m\times n}$ such that
$\boldsymbol{M}_{ij}$'s are i.i.d. standard Gaussian random variables.
For any constant $0<\xi<\sqrt{n}-\sqrt{m}$,
\begin{equation}
\sigma_{\min}\left(\boldsymbol{M}\boldsymbol{M}^{*}\right)>\left(\sqrt{n}-\sqrt{m}-\xi\right)^{2}\label{eq:sigma_min}
\end{equation}
with probability at least\footnote{Note that this follows from \cite[Proposition 5.34 and Theorem 5.32]{Vershynin2012}
by observing that $\sigma_{\min}\left(\boldsymbol{M}\right)$ is a
1-Lipschitz function.} $1-\exp\left(-\xi^{2}/2\right)$.\end{lem}

Lemma \ref{lemma-Concentration-of-Norm-M} indicates that if $n>\frac{2}{1-\sqrt{\alpha}}$,
then the event
\[
\lambda_{\min}\left(\frac{1}{n}\boldsymbol{A}\boldsymbol{A}^{\top}\right)\geq\left(1-\sqrt{\alpha}-\frac{2}{n}\right)^{2}
\]
occurs with probability at least $1-\exp\left(-\frac{2}{n}\right)$.
Conditioned on this event, we have
\begin{equation}
\det\left(\frac{1}{n}\boldsymbol{A}\boldsymbol{A}^{\top}\right)=\text{det}^{\delta}\left(\frac{1}{n}\boldsymbol{A}\boldsymbol{A}^{\top}\right)
\end{equation}
holds for the numerical value $\delta=\left(1-\sqrt{\alpha}-\frac{2}{n}\right)^{2}$.
This taken collectively with (\ref{eq:GaussianDetAA-1}) (by setting
$\tau=\frac{1}{4}$) yields that for any $n>\max\left\{ \frac{2}{1-\sqrt{\alpha}},5\right\} $,

\begin{align*}
\det\left(\frac{1}{n}\boldsymbol{A}\boldsymbol{A}^{\top}\right) & =\text{det}^{\delta}\left(\frac{1}{n}\boldsymbol{A}\boldsymbol{A}^{\top}\right)\quad\\
\text{and}\quad\text{det}^{\delta}\left(\frac{1}{n}\boldsymbol{A}\boldsymbol{A}^{\top}\right) & >\frac{1}{e^{\sqrt{\frac{4\alpha}{\delta n}}}}e^{\mathbb{E}\left[\log\text{det}^{\delta}\left(\frac{1}{n}\boldsymbol{A}\boldsymbol{A}^{\top}\right)\right]}
\end{align*}
hold simultaneously with probability at least $1-\exp\left(-\frac{2}{n}\right)-2\exp\left(-\frac{n}{2}\right)>1/n$.
Since $\det\left(\frac{1}{n}\boldsymbol{A}\boldsymbol{A}^{\top}\right)$
is non-negative, taking expectation gives
\begin{equation}
\mathbb{E}\left[\det\left(\frac{1}{n}\boldsymbol{A}\boldsymbol{A}^{\top}\right)\right]\geq\frac{1}{n}\cdot\frac{1}{e^{\sqrt{\frac{4\alpha}{\delta n}}}}e^{\mathbb{E}\left[\log\text{det}^{\delta}\left(\frac{1}{n}\boldsymbol{A}\boldsymbol{A}^{\top}\right)\right]},
\end{equation}
and therefore for any $n>\max\left\{ \frac{2}{1-\sqrt{\alpha}},7\right\} $
and $\delta=\left(1-\sqrt{\alpha}-\frac{2}{n}\right)^{2}$,
\begin{align*}
 & \frac{1}{n}\mathbb{E}\left[\log\text{det}^{\delta}\left(\frac{1}{n}\boldsymbol{A}\boldsymbol{A}^{\top}\right)\right]\\
 & \leq\frac{1}{n}\log\mathbb{E}\left[\det\left(\frac{1}{n}\boldsymbol{A}\boldsymbol{A}^{\top}\right)\right]+\frac{\log n}{n}+\frac{\sqrt{\frac{4\alpha}{\delta}}}{n^{1.5}}\\
 & \leq\left(1-\alpha\right)\log\frac{1}{1-\alpha}-\alpha+\frac{\log\left(e^{2}m\right)+2\log n}{2n}+\frac{2\sqrt{\alpha}n^{-1.5}}{1-\sqrt{\alpha}-\frac{2}{n}}\\
 & \leq\left(1-\alpha\right)\log\frac{1}{1-\alpha}-\alpha+\frac{2\log n}{n}+\frac{2\sqrt{\alpha}}{\left(1-\sqrt{\alpha}-\frac{2}{n}\right)}\frac{1}{n^{1.5}},
\end{align*}
where the second inequality follows from (\ref{eq:UB_logEdet}). Putting
this and (\ref{eq:GaussianDetAA-1}) together gives that for any $n>\max\left\{ \frac{2}{1-\sqrt{\alpha}},7,\frac{2}{\sqrt{\tau}}\right\} $,
\begin{align*}
 & \frac{1}{n}\log\det\left(\frac{1}{n}\boldsymbol{A}\boldsymbol{A}^{\top}\right)\leq\frac{4\sqrt{\alpha}}{\left(1-\sqrt{\alpha}-\frac{2}{n}\right)}\sqrt{\frac{\tau}{n}}-\alpha\\
 & \quad\quad\text{ }+\left(1-\alpha\right)\log\frac{1}{1-\alpha}+\frac{2\log n}{n}+\frac{2\sqrt{\alpha}}{1-\sqrt{\alpha}-\frac{2}{n}}\frac{1}{n^{1.5}}\\
 & \quad<\left(1-\alpha\right)\log\frac{1}{1-\alpha}-\alpha+\frac{2\log n}{n}\\
 & \quad\quad\quad+\frac{4\sqrt{\alpha}}{\left(1-\sqrt{\alpha}-\frac{2}{n}\right)\sqrt{n}}\left(\sqrt{\tau}+\frac{1}{2n}\right)\\
 & \quad\leq\left(1-\alpha\right)\log\frac{1}{1-\alpha}-\alpha+\frac{2\log n}{n}+\frac{5\sqrt{\alpha}}{1-\sqrt{\alpha}-\frac{2}{n}}\sqrt{\frac{\tau}{n}}
\end{align*}
with probability exceeding $1-2\exp\left(-2\tau n\right)$, as claimed.

(2) In order to derive a lower bound on $\log\det\left(\frac{1}{n}\boldsymbol{A}\boldsymbol{A}^{\top}\right)$
based on (\ref{eq:GaussianDetAA-1}), one would first need to bound
$\mathbb{E}\left[\log\text{det}^{\delta}\left(\frac{1}{n}\boldsymbol{A}\boldsymbol{A}^{\top}\right)\right]$
from below. Observe the following consequence from (\ref{eq:UB_Z}):
\begin{align}
 & \mathbb{E}\left[e^{Z}\right]\leq\mathbb{E}\left[e^{\left|Z\right|}\right]\nonumber \\
 & \text{ }=-e^{z}\mathbb{P}\left(\left|Z\right|>z\right)\mid_{z=0}^{\infty}+\int_{0}^{\infty}e^{z}\mathbb{P}\left(\left|Z\right|>z\right)\mathrm{d}z\\
 & \text{ }\leq1+\int_{0}^{\infty}2\exp\left(z-\frac{\delta z^{2}}{8\alpha}\right)\mathrm{d}z\nonumber \\
 & \text{ }<1+4\sqrt{\frac{2\pi\alpha}{\delta}}\exp\left(\frac{2\alpha}{\delta}\right).\label{eq:UB_Ee_Z}
\end{align}
Taking the logarithm on both sides of (\ref{eq:UB_Ee_Z}) and plugging
in the expression of $Z$ yields
\begin{align*}
\log\mathbb{E}\left[\text{det}^{\delta}\left(\frac{1}{n}\boldsymbol{A}\boldsymbol{A}^{\top}\right)\right] & \leq\mathbb{E}\left[\log\text{det}^{\delta}\left(\frac{1}{n}\boldsymbol{A}\boldsymbol{A}^{\top}\right)\right]\\
 & +\log\left[1+4\sqrt{\frac{2\pi\alpha}{\delta}}\exp\left(\frac{2\alpha}{\delta}\right)\right],
\end{align*}
leading to
\begin{align}
 & \frac{1}{n}\mathbb{E}\left[\log\text{det}^{\delta}\left(\frac{1}{n}\boldsymbol{A}\boldsymbol{A}^{\top}\right)\right]\nonumber \\
 & \quad\geq\frac{1}{n}\log\mathbb{E}\left[\text{det}^{\delta}\left(\frac{1}{n}\boldsymbol{A}\boldsymbol{A}^{\top}\right)\right]-\frac{\log\left[1+4\sqrt{\frac{2\pi\alpha}{\delta}}\exp\left(\frac{2\alpha}{\delta}\right)\right]}{n}\nonumber \\
 & \quad\geq\frac{1}{n}\log\mathbb{E}\left[\text{det}\left(\frac{1}{n}\boldsymbol{A}\boldsymbol{A}^{\top}\right)\right]-\frac{\log\left[1+4\sqrt{\frac{2\pi\alpha}{\delta}}\exp\left(\frac{2\alpha}{\delta}\right)\right]}{n}\nonumber \\
 & \quad\geq\frac{1}{n}\log\mathbb{E}\left[\text{det}\left(\frac{1}{n}\boldsymbol{A}\boldsymbol{A}^{\top}\right)\right]-\frac{5\alpha}{n\delta}.\label{eq:LowerBoundELogDetEpsilon}
\end{align}
for any $\delta\leq\alpha$. Here, the last inequality follows from
simple numerical inequality $1+4\sqrt{2\pi x}\exp(2x)\leq\exp(5x)$
for any $x\geq1$. Consequently, for any $\delta\leq\alpha$ 
\begin{align}
\frac{1}{n}\mathbb{E}\left[\log\text{det}^{\delta}\left(\frac{1}{n}\boldsymbol{A}\boldsymbol{A}^{\top}\right)\right] & \geq\left(1-\alpha\right)\log\frac{1}{1-\alpha}-\alpha\nonumber \\
 & \quad\quad-\frac{\log\left(n+1\right)}{n}-\frac{5\alpha}{n\delta}.\label{eq:LowerBoundELogDetEpsilon-Gaussian}
\end{align}
This together with (\ref{eq:GaussianDetAA-1}) characterizes a lower
bound on $\log\det^{\delta}\left(\frac{1}{n}\boldsymbol{A}\boldsymbol{A}^{\top}\right)$. 

It remains to quantify the gap between $\log\det^{\delta}\left(\frac{1}{n}\boldsymbol{A}\boldsymbol{A}^{\top}\right)$
and $\log\det\left(\frac{1}{n}\boldsymbol{A}\boldsymbol{A}^{\top}\right)$.
On the one hand, the inequality (\ref{eq:CardinalityEpsilon}) indicates
that for any $n\geq\left(\frac{\alpha}{1-\alpha-\frac{1}{n}}+4\sqrt{\alpha\tau}\right)^{3}$,
\begin{align}
 & \mathbb{P}\left\{ \lambda_{\max}\left(\boldsymbol{A}\boldsymbol{A}^{\top}\right)<\frac{1}{n^{1/3}}\right\} \nonumber \\
 & \text{ }=\mathbb{P}\left\{ \frac{\text{card}\left\{ i\mid\lambda_{i}\left(\frac{1}{n}\boldsymbol{A}\boldsymbol{A}^{\top}\right)<\frac{1}{n^{1/3}}\right\} }{n}>1\right\} \nonumber \\
 & \text{ }\leq\mathbb{P}\left\{ \frac{\text{card}\left\{ i\mid\lambda_{i}\left(\frac{1}{n}\boldsymbol{A}\boldsymbol{A}^{\top}\right)<\frac{1}{n^{1/3}}\right\} }{n}>\frac{\frac{\alpha}{1-\alpha-\frac{1}{n}}+4\sqrt{\alpha\tau}}{n^{1/3}}\right\} \nonumber \\
 & \text{ }\leq2e^{-\tau n}.
\end{align}
On the other hand, it follows from \cite[Theorem 4.5]{ChenDongarra2005}
that for any $n\geq\frac{6.414}{1-\alpha}\cdot e^{\frac{\tau}{1-\alpha}}$,
\begin{align}
 & \mathbb{P}\left\{ \frac{\lambda_{\max}\left(\boldsymbol{A}\boldsymbol{A}^{\top}\right)}{\lambda_{\min}\left(\boldsymbol{A}\boldsymbol{A}^{\top}\right)}>n^{2}\right\} \nonumber \\
 & \text{ }=\text{}\mathbb{P}\left\{ \frac{\lambda_{\max}\left(\boldsymbol{A}\boldsymbol{A}^{\top}\right)}{\lambda_{\min}\left(\boldsymbol{A}\boldsymbol{A}^{\top}\right)}>\frac{n^{2}}{\left(n-m+1\right)^{2}}\cdot\left(n-m+1\right)^{2}\right\} \nonumber \\
 & \text{ }\leq\text{}\frac{1}{\sqrt{2\pi}}\left(\frac{6.414}{(1-\alpha)n}\right)^{\left(1-\alpha\right)n}\leq\frac{1}{\sqrt{2\pi}}e^{-\tau n}.
\end{align}
The above two probability bounds taken collectively imply that for
any $n\geq\max\left\{ \frac{6.414}{1-\alpha}\cdot e^{\frac{\tau}{1-\alpha}},\left(\frac{\alpha}{1-\alpha-\frac{1}{n}}+4\sqrt{\alpha\tau}\right)^{3}\right\} $,
\begin{align}
 & \small\mathbb{P}\left\{ \lambda_{\min}\left(\frac{1}{n}\boldsymbol{A}\boldsymbol{A}^{\top}\right)<\frac{1}{n^{7/3}}\right\} \leq\text{ }\mathbb{P}\left\{ \lambda_{\max}\left(\boldsymbol{A}\boldsymbol{A}^{\top}\right)<\frac{1}{n^{1/3}}\right\} \nonumber \\
 & \quad\small+\mathbb{P}\left\{ \lambda_{\max}\left(\boldsymbol{A}\boldsymbol{A}^{\top}\right)\geq\frac{1}{n^{1/3}}\text{ and }\lambda_{\min}\left(\frac{1}{n}\boldsymbol{A}\boldsymbol{A}^{\top}\right)<\frac{1}{n^{7/3}}\right\} \nonumber \\
 & \quad\leq\text{ }\small\mathbb{P}\left\{ \lambda_{\max}\left(\boldsymbol{A}\boldsymbol{A}^{\top}\right)<\frac{1}{n^{1/3}}\right\} +\mathbb{P}\left\{ \frac{\lambda_{\max}\left(\boldsymbol{A}\boldsymbol{A}^{\top}\right)}{\lambda_{\min}\left(\boldsymbol{A}\boldsymbol{A}^{\top}\right)}>n^{2}\right\} \nonumber \\
 & \quad\leq\text{ }2e^{-\tau n}+\frac{1}{\sqrt{2\pi}}e^{-\tau n}<3e^{-\tau n}.\label{eq:Lambda_min_AA}
\end{align}

Consequently, by setting $\delta=n^{-1/3}$ and applying Lemma \ref{lem:GaussianLocalSpectrum}
one obtains 
\begin{align*}
 & \frac{1}{n}\log\det\left(\frac{1}{n}\boldsymbol{A}\boldsymbol{A}^{\top}\right)\geq\frac{1}{n}\log\mathrm{det}^{\delta}\left(\frac{1}{n}\boldsymbol{A}\boldsymbol{A}^{\top}\right)\\
 & \quad\quad-\sum_{i:\lambda_{i}\left(\frac{1}{n}\boldsymbol{A}\boldsymbol{A}^{\top}\right)<\frac{1}{n^{1/3}}}\log\frac{1}{n^{1/3}}-\log\lambda_{\min}\left(\frac{1}{n}\boldsymbol{A}\boldsymbol{A}^{\top}\right)\\
 & \text{ }\geq\frac{1}{n}\log\mathrm{det}^{\delta}\left(\frac{1}{n}\boldsymbol{A}\boldsymbol{A}^{\top}\right)-\frac{\footnotesize\text{card}\left\{ i\mid\lambda_{i}\left(\frac{1}{n}\boldsymbol{A}\boldsymbol{A}^{\top}\right)<\frac{1}{n^{\frac{1}{3}}}\right\} \log\left(n^{2}\right)}{n}\\
 & \text{ }>\small\frac{1}{n}\log\mathrm{det}^{\delta}\left(\frac{1}{n}\boldsymbol{A}\boldsymbol{A}^{\top}\right)-\left(\frac{2\alpha}{1-\alpha-\frac{1}{n}}+8\sqrt{\alpha\tau}\right)\cdot\frac{\log n}{n^{1/3}}
\end{align*}
with probability exceeding $1-5\exp\left(-\tau n\right)$. By making
use of (\ref{eq:GaussianDetAA-1}), one obtains that when $\delta=n^{-1/3}$,
\begin{align}
 & \footnotesize\mathbb{P}\left\{ \left|\frac{1}{n}\log\mathrm{det}^{\delta}\left(\frac{1}{n}\boldsymbol{A}\boldsymbol{A}^{\top}\right)-\mathbb{E}\left[\frac{1}{n}\log\mathrm{det}^{\delta}\left(\frac{1}{n}\boldsymbol{A}\boldsymbol{A}^{\top}\right)\right]\right|>\frac{\sqrt{8\tau\alpha}}{n^{1/3}}\right\} \nonumber \\
 & \quad<2\exp\left(-\tau n\right).
\end{align}
Putting the above two bounds together implies that for any $n>\max\left\{ \frac{6.414}{1-\alpha}\cdot e^{\frac{\tau}{1-\alpha}},\left(\frac{\alpha}{1-\alpha-\frac{1}{n}}+4\sqrt{\alpha\tau}\right)^{3}\right\} $,
\begin{align}
 & \frac{1}{n}\log\det\left(\frac{1}{n}\boldsymbol{A}\boldsymbol{A}^{\top}\right)\nonumber \\
 & \geq\frac{1}{n}\log\mathrm{det}^{\delta}\left(\frac{1}{n}\boldsymbol{A}\boldsymbol{A}^{\top}\right)-\left(\frac{2\alpha}{1-\alpha-\frac{1}{n}}+8\sqrt{\alpha\tau}\right)\frac{\log n}{n^{1/3}}\nonumber \\
 & \geq\footnotesize\mathbb{E}\left[\frac{1}{n}\log\mathrm{det}^{\delta}\left(\frac{1}{n}\boldsymbol{A}\boldsymbol{A}^{\top}\right)\right]-\frac{\sqrt{8\tau\alpha}}{n^{\frac{1}{3}}}-\left(\frac{2\alpha}{1-\alpha-\frac{1}{n}}+8\sqrt{\alpha\tau}\right)\frac{\log n}{n^{\frac{1}{3}}}\nonumber \\
 & >\left(1-\alpha\right)\log\frac{1}{1-\alpha}-\alpha-\frac{\log\left(n+1\right)}{n}-\frac{5\alpha}{n^{2/3}}\nonumber \\
 & \text{ }\quad\quad\quad\quad\quad\small-\left(\frac{2\alpha}{1-\alpha-\frac{1}{n}}+10\sqrt{\alpha\tau}\right)\cdot\frac{\log n}{n^{1/3}}\label{eq:last_inequality}\\
 & >\left(1-\alpha\right)\log\frac{1}{1-\alpha}-\alpha-\left(\frac{2}{1-\alpha-\frac{1}{n}}+10\sqrt{\tau}\right)\cdot\frac{\log n}{n^{1/3}},\nonumber 
\end{align}
with probability exceeding $1-7\exp(-\tau n)$. Here, (\ref{eq:last_inequality})
follows from (\ref{eq:LowerBoundELogDetEpsilon-Gaussian}), and the
last inequality makes use of the fact that $\frac{5}{n^{2/3}}+\frac{\log\left(n+1\right)}{n}\leq\frac{2\log n}{n^{1/3}}$
for all $n>6$.

\section{Proof of Lemma \ref{lemma-Expected-Log-Determinant-LambdaStatistics}
\label{sec:Proof-of-lemma-Expected-Log-Determinant-LambdaStatistics}}

Suppose that the singular value decomposition of the real-valued $\boldsymbol{A}$
is given by $\boldsymbol{A}=\boldsymbol{U}_{\boldsymbol{A}}\left[\begin{array}{c}
\boldsymbol{\Sigma}_{\boldsymbol{A}}\\
{\bf 0}
\end{array}\right]\boldsymbol{V}_{\boldsymbol{A}}^{\top}$, where $\boldsymbol{\Sigma}_{\boldsymbol{A}}$ is a diagonal matrix
containing all $k$ singular values of $\boldsymbol{A}$. One can
then write 
\begin{align}
 & \log\det\left(\epsilon\boldsymbol{I}+\boldsymbol{A}^{\top}\boldsymbol{B}^{-1}\boldsymbol{A}\right)\nonumber \\
 & \text{ }=\log\det\left(\epsilon\boldsymbol{I}+\boldsymbol{V}_{\boldsymbol{A}}\left[\boldsymbol{\Sigma}_{\boldsymbol{A}}\text{ }{\bf 0}\right]\boldsymbol{U}_{\boldsymbol{A}}^{\top}\boldsymbol{B}^{-1}\boldsymbol{U}_{\boldsymbol{A}}\left[\begin{array}{c}
\boldsymbol{\Sigma}_{\boldsymbol{A}}\\
{\bf 0}
\end{array}\right]\boldsymbol{V}_{\boldsymbol{A}}^{\top}\right)\nonumber \\
 & \text{ }=\log\det\left(\epsilon\boldsymbol{I}+\boldsymbol{\Sigma}_{\boldsymbol{A}}\left(\tilde{\boldsymbol{B}}^{-1}\right)_{[k]}\boldsymbol{\Sigma}_{\boldsymbol{A}}\right)\nonumber \\
 & \text{ }\geq\log\det\left(\frac{1}{n}\boldsymbol{\Sigma}_{\boldsymbol{A}}^{2}\right)-\log\det\left\{ \frac{1}{n}\left(\tilde{\boldsymbol{B}}^{-1}\right)_{[k]}\right\} \label{eq:SeparationLogDetI_Plus_ABA}
\end{align}
where $\tilde{\boldsymbol{B}}=\boldsymbol{U}_{\boldsymbol{A}}^{\top}\boldsymbol{B}\boldsymbol{U}_{\boldsymbol{A}}\sim\mathcal{W}_{m}\left(n-k,\boldsymbol{U}_{\boldsymbol{A}}^{\top}\boldsymbol{U}_{\boldsymbol{A}}\right)=\mathcal{W}_{m}\left(n-k,\boldsymbol{I}_{m}\right)$
from the property of Wishart distribution. Here, $\left(\tilde{\boldsymbol{B}}^{-1}\right)_{[k]}$
denotes the leading $k\times k$ minor consisting of matrix elements
of $\tilde{\boldsymbol{B}}^{-1}$ in rows and columns from $1$ to
$k$, which is independent of $\boldsymbol{A}$ by Gaussianality. 

Note that $\frac{1}{n}\log\det\left(\frac{1}{n}\boldsymbol{\Sigma}_{\boldsymbol{A}}^{2}\right)=\frac{1}{n}\log\det\left(\frac{1}{n}\boldsymbol{A}^{\top}\boldsymbol{A}\right)$.
Then Lemma \ref{lem:GaussianLogDet_LB} implies that for any $\tau>0$
and sufficiently large $n$,
\begin{align}
 & \frac{1}{n}\log\det\left(\frac{1}{n}\boldsymbol{\Sigma}_{\boldsymbol{A}}^{2}\right)=\frac{1}{n}\log\det\left(\frac{1}{n}\boldsymbol{A}^{\top}\boldsymbol{A}\right)\nonumber \\
 & \text{ }=\frac{1}{n}\log\det\left(\frac{m}{n}\boldsymbol{I}_{k}\right)+\frac{m}{n}\frac{1}{m}\log\det\left(\frac{1}{m}\boldsymbol{A}^{\top}\boldsymbol{A}\right)\nonumber \\
 & \text{ }\geq\text{ }\beta\log\alpha+\alpha\left(-\left(1-\frac{\beta}{\alpha}\right)\log\left(1-\frac{\beta}{\alpha}\right)-\frac{\beta}{\alpha}\right)-\hat{\kappa}\nonumber \\
 & \text{ }\geq-\left(\alpha-\beta\right)\log\left(\frac{\alpha-\beta}{\alpha}\right)-\beta+\beta\log\alpha-\hat{\kappa}\nonumber \\
 & \text{ }=-\left(\alpha-\beta\right)\log\left(\alpha-\beta\right)-\beta+\alpha\log\alpha-\hat{\kappa},\label{eq:ConcentrationLogDetAAT}
\end{align}
with probability exceeding $1-7\exp\left(-\tau^{2}n\right)$, where
\begin{equation}
\hat{\kappa}:=\frac{\left(\frac{2\alpha}{1-\frac{\beta}{\alpha}-\frac{1}{n}}+10\alpha\tau\right)\log m}{m^{1/3}}\leq\frac{\left(\frac{2}{1-\frac{\beta}{\alpha}-\frac{1}{n}}+10\tau\right)\log n}{n^{1/3}}.
\end{equation}
 On the other hand, it is well known (e.g. \cite[Theorem 2.3.3]{Fujikoshi2010})
that for a Wishart matrix $\tilde{\boldsymbol{B}}\sim\mathcal{W}_{m}\left(n-k,\boldsymbol{I}_{m}\right)$,
$\left(\tilde{\boldsymbol{B}}^{-1}\right)_{[k]}^{-1}$ also follows
the Wishart distribution, that is,
\begin{equation}
\left(\tilde{\boldsymbol{B}}^{-1}\right)_{[k]}^{-1}\sim\mathcal{W}_{k}\left(n-m,\boldsymbol{I}_{k}\right).
\end{equation}
By setting $\zeta:=\max\left\{ \frac{\beta}{\alpha},\frac{\beta}{1-\alpha}\right\} ,$then
one can obtain from Lemma \ref{lem:GaussianLogDet_LB} that for sufficiently
large $n$,
\begin{align}
 & \frac{1}{n}\log\det\left(\frac{1}{n}\left(\tilde{\boldsymbol{B}}^{-1}\right)_{[k]}^{-1}\right)\nonumber \\
 & \text{ }=\frac{1}{n}\log\det\left(\frac{1}{n-m}\left(\tilde{\boldsymbol{B}}^{-1}\right)_{[k]}^{-1}\right)+\frac{1}{n}\log\det\left(\frac{n-m}{n}\boldsymbol{I}_{k}\right)\nonumber \\
 & \text{ }\leq\left(1-\alpha\right)\left\{ -\left(1-\frac{\beta}{1-\alpha}\right)\log\left(1-\frac{\beta}{1-\alpha}\right)-\frac{\beta}{1-\alpha}\right\} \nonumber \\
 & \quad\quad\quad\quad+\beta\log\left(1-\alpha\right)+\frac{2\log n}{n}+\frac{\frac{5\sqrt{\zeta}}{\left(1-\sqrt{\zeta}-\frac{2}{n}\right)}\tau}{\sqrt{n}}\nonumber \\
 & \text{ }=-\left(1-\alpha-\beta\right)\log\left(1-\frac{\beta}{1-\alpha}\right)-\beta+\beta\log\left(1-\alpha\right)\nonumber \\
 & \quad\quad\quad\quad+\frac{2\log n}{n}+\frac{\frac{5\sqrt{\zeta}}{\left(1-\sqrt{\zeta}-\frac{2}{n}\right)}\tau}{\sqrt{n}}\label{eq:ConcentrationLogDetBMinor}
\end{align}
holds with probability exceeding $1-2\exp\left(-2\tau^{2}n\right)$.

Combining (\ref{eq:SeparationLogDetI_Plus_ABA}), (\ref{eq:ConcentrationLogDetAAT})
and (\ref{eq:SeparationLogDetI_Plus_ABA}) suggests that for any $\tau>0$
and sufficiently large $n$, one has
\begin{align*}
 & \frac{1}{n}\log\det\left(\epsilon\boldsymbol{I}_{k}+\boldsymbol{A}^{\top}\boldsymbol{B}^{-1}\boldsymbol{A}\right)\\
 & \text{ }\geq-\left(\alpha-\beta\right)\log\left(\alpha-\beta\right)+\alpha\log\alpha-\beta\log\left(1-\alpha\right)+\\
 & \text{ }\left(1-\alpha-\beta\right)\log\frac{1-\alpha-\beta}{1-\alpha}-\frac{\left(\frac{2}{1-\zeta-\frac{1}{n}}+\frac{10-5\sqrt{\zeta}}{\left(1-\sqrt{\zeta}-\frac{2}{n}\right)}\tau\right)\log n}{n^{1/3}}
\end{align*}
with probability exceeding $1-9\exp\left(-\tau^{2}n\right)$.

\bibliographystyle{IEEEtran}

\bibliographystyle{IEEEtran}
\bibliography{bibfileMinimax}

\begin{IEEEbiographynophoto}{Yuxin Chen} (S'09) received the B.S. in Microelectronics with High Distinction from Tsinghua University in 2008, the M.S. in Electrical and Computer Engineering from the University of Texas at Austin in 2010, the M.S. in Statistics from Stanford University in 2013, and the Ph.D. in Electrical Engineering from Stanford University in 2015. He joins the Department of Electrical Engineering at Princeton University as an assistant professor in February 2017. His research interests include mathematical optimization, information theory, high-dimensional statistics, and statistical learning. \end{IEEEbiographynophoto} 

\begin{IEEEbiographynophoto}{Andrea J. Goldsmith} (S'90-M'93-SM'99-F'05)
is the Stephen Harris professor in the School of Engineering and a professor of Electrical Engineering at Stanford University. She was previously on the faculty of Electrical Engineering at Caltech. Her research interests are in information theory and communication theory, and their application to wireless communications and related fields. She co-founded and serves as Chief Scientist of Accelera, Inc., and previously co-founded and served as CTO of Quantenna Communications, Inc. She has also held industry positions at Maxim Technologies, Memorylink Corporation, and AT\&T Bell Laboratories. Dr. Goldsmith is a Fellow of the IEEE and of Stanford, and she has received several awards for her work, including the IEEE Communications Society and Information Theory Society joint paper award, the IEEE Communications Society Best Tutorial Paper Award, the National Academy of Engineering Gilbreth Lecture Award, the IEEE ComSoc Communications Theory Technical Achievement Award, the IEEE ComSoc Wireless Communications Technical Achievement Award, the Alfred P. Sloan Fellowship, and the Silicon Valley/San Jose Business Journal's Women of Influence Award. She is author of the book ``Wireless Communications'' and co-author of the books ``MIMO Wireless Communications'' and ``Principles of Cognitive Radio,'' all published by Cambridge University Press. She received the B.S., M.S. and Ph.D. degrees in Electrical Engineering from U.C. Berkeley.  

Dr. Goldsmith has served on the Steering Committee for the IEEE Transactions on Wireless Communications and as editor for the IEEE Transactions on Information Theory, the Journal on Foundations and Trends in Communications and Information Theory and in Networks, the IEEE Transactions on Communications, and the IEEE Wireless Communications Magazine. She participates actively in committees and conference organization for the IEEE Information Theory and Communications Societies and has served on the Board of Governors for both societies. She has also been a Distinguished Lecturer for both societies, served as President of the IEEE Information Theory Society in 2009, founded and chaired the student committee of the IEEE Information Theory society, and chaired the Emerging Technology Committee of the IEEE Communications Society. At Stanford she received the inaugural University Postdoc Mentoring Award, served as Chair of Stanfords Faculty Senate in 2009 and currently serves on its Faculty Senate and on its Budget Group.
\end{IEEEbiographynophoto} \begin{IEEEbiographynophoto}{Yonina C. Eldar} (S'98-M'02-SM'07-F'12) received the B.Sc. degree in physics and  the B.Sc. degree in electrical engineering both from Tel-Aviv University (TAU), Tel-Aviv,  Israel, in 1995 and 1996, respectively, and the Ph.D. degree in electrical engineering and computer science from the  Massachusetts Institute of Technology (MIT), Cambridge, in 2002.

 From January 2002 to July 2002, she was a Postdoctoral Fellow at the  Digital Signal Processing Group at MIT. She is currently a Professor in the Department of Electrical  Engineering at the Technion-Israel Institute of Technology, Haifa and holds the  The Edwards Chair in Engineering. She is  also a Research Affiliate with the Research Laboratory of Electronics at MIT  and a Visiting Professor at Stanford University, Stanford, CA. Her research interests are in the broad areas of  statistical signal processing, sampling theory and compressed sensing,  optimization methods, and their applications to biology and optics.

Dr. Eldar was in the program for outstanding students at TAU from 1992 to 1996. In 1998, she held the Rosenblith Fellowship for study in electrical engineering at MIT, and in 2000, she held an IBM Research Fellowship. From  2002 to 2005, she was a Horev Fellow of the Leaders in Science and  Technology program at the Technion and an Alon Fellow. In 2004, she was  awarded the Wolf Foundation Krill Prize for Excellence in Scientific  Research, in 2005 the Andre and Bella Meyer Lectureship, in 2007 the Henry  Taub Prize for Excellence in Research, in 2008 the Hershel Rich Innovation  Award, the Award for Women with Distinguished Contributions, the Muriel \& David Jacknow Award for Excellence in Teaching, and the Technion Outstanding  Lecture Award, in 2009 the Technion's Award for Excellence in Teaching, in 2010 the Michael  Bruno Memorial Award from the Rothschild Foundation, and in 2011 the Weizmann Prize for Exact Sciences.   In 2012 she was elected to the Young Israel Academy of Science and to the Israel Committee for Higher Education, and elected an IEEE Fellow.  In 2013 she received the Technion's Award for Excellence in Teaching, the Hershel Rich Innovation Award, and the IEEE Signal Processing Technical Achievement Award. She received several best paper awards together with her research students and colleagues.  She received several best paper awards together with her research students and colleagues. She is the Editor in Chief of Foundations and Trends in Signal Processing. In the past, she was a Signal Processing Society Distinguished Lecturer, member of the IEEE Signal Processing Theory and Methods and Bio Imaging Signal Processing technical committees, and served as an associate editor for the IEEE Transactions On Signal Processing, the EURASIP Journal of Signal Processing, the SIAM Journal on Matrix Analysis and Applications, and the SIAM Journal on Imaging Sciences.
\end{IEEEbiographynophoto} 
\end{document}